%
%
%
%
%
%
%
%
\def\standardrisposta{s }\def\reducedrisposta{r }
\def\mplarisposta{mpla }\def\zerorisposta{z }
\def\doublerisposta{d }\def\cartarisposta{e }\def\amsrisposta{y }
\newcount\ingrandimento \newcount\sinnota \newcount\dimnota
\newcount\unoduecol \newdimen\collhsize \newdimen\tothsize
\newdimen\fullhsize \newcount\controllorisposta \sinnota=1
\newskip\infralinea  \global\controllorisposta=0
%
%
%
%
%
\def\risposta{s }
\def\srisposta{e }
\def\arisposta{y }
\ifx\risposta\standardrisposta \ingrandimento=1200
\message {>> This will come out UNREDUCED << }
\dimnota=2 \unoduecol=1 \global\controllorisposta=1 \fi
\ifx\risposta\reducedrisposta \ingrandimento=1095 \dimnota=1
\unoduecol=1  \global\controllorisposta=1
\message {>> This will come out REDUCED << } \fi
\ifx\risposta\doublerisposta \ingrandimento=1000 \dimnota=2
\unoduecol=2   \message {>> You must print this in
LANDSCAPE orientation << } \global\controllorisposta=1 \fi
\ifx\risposta\mplarisposta \ingrandimento=1000 \dimnota=1
\message {>> Mod. Phys. Lett. A format << }
\unoduecol=1 \global\controllorisposta=1 \fi
\ifx\risposta\zerorisposta \ingrandimento=1000 \dimnota=2
\message {>> Zero Magnification format << }
\unoduecol=1 \global\controllorisposta=1 \fi
\ifnum\controllorisposta=0  \ingrandimento=1200
\message {>>> ERROR IN INPUT, I ASSUME STANDARD
UNREDUCED FORMAT <<< }  \dimnota=2 \unoduecol=1 \fi
\magnification=\ingrandimento
%
%
%
%
\newdimen\eucolumnsize \newdimen\eudoublehsize \newdimen\eudoublevsize
\newdimen\uscolumnsize \newdimen\usdoublehsize \newdimen\usdoublevsize
\newdimen\eusinglehsize \newdimen\eusinglevsize \newdimen\ussinglehsize
\newskip\standardbaselineskip \newdimen\ussinglevsize
\newskip\reducedbaselineskip \newskip\doublebaselineskip
\eucolumnsize=12.0truecm    
\eudoublehsize=25.5truecm   
\eudoublevsize=6.5truein    
\uscolumnsize=4.4truein     
\usdoublehsize=9.4truein    
\usdoublevsize=6.8truein    
\eusinglehsize=6.5truein    
\eusinglevsize=24truecm     
\ussinglehsize=6.5truein    
\ussinglevsize=8.9truein    
\standardbaselineskip=16pt plus.2pt  
\reducedbaselineskip=14pt plus.2pt   
\doublebaselineskip=12pt plus.2pt    
%
%
\def\Portoffset{}
\def\Landoffset{}
\ifx\risposta\mplarisposta \def\Portoffset{\hoffset=1.8truecm} \fi
%
%
\def\Landspec{}
\tolerance=10000
\parskip=0pt plus2pt  \leftskip=0pt \rightskip=0pt
%
%
\ifx\risposta\standardrisposta \infralinea=\standardbaselineskip \fi
\ifx\risposta\reducedrisposta  \infralinea=\reducedbaselineskip \fi
\ifx\risposta\doublerisposta   \infralinea=\doublebaselineskip \fi
\ifx\risposta\mplarisposta     \infralinea=13pt \fi
\ifx\risposta\zerorisposta     \infralinea=12pt plus.2pt\fi
\ifnum\controllorisposta=0    \infralinea=\standardbaselineskip \fi
\ifx\risposta\doublerisposta   \Landoffset \else \Portoffset \fi
\ifx\risposta\doublerisposta \ifx\srisposta\cartarisposta
\tothsize=\eudoublehsize \collhsize=\eucolumnsize
\vsize=\eudoublevsize  \else  \tothsize=\usdoublehsize
\collhsize=\uscolumnsize \vsize=\usdoublevsize \fi \else
\ifx\srisposta\cartarisposta \tothsize=\eusinglehsize
\vsize=\eusinglevsize \else  \tothsize=\ussinglehsize
\vsize=\ussinglevsize \fi \collhsize=4.4truein \fi
\ifx\risposta\mplarisposta \tothsize=5.0truein
\vsize=7.8truein \collhsize=4.4truein \fi
%
%
%
%
\newcount\contaeuler \newcount\contacyrill \newcount\contaams
\font\ninerm=cmr9  \font\eightrm=cmr8  \font\sixrm=cmr6
\font\ninei=cmmi9  \font\eighti=cmmi8  \font\sixi=cmmi6
\font\ninesy=cmsy9  \font\eightsy=cmsy8  \font\sixsy=cmsy6
\font\ninebf=cmbx9  \font\eightbf=cmbx8  \font\sixbf=cmbx6
\font\ninett=cmtt9  \font\eighttt=cmtt8  \font\nineit=cmti9
\font\eightit=cmti8 \font\ninesl=cmsl9  \font\eightsl=cmsl8
\skewchar\ninei='177 \skewchar\eighti='177 \skewchar\sixi='177
\skewchar\ninesy='60 \skewchar\eightsy='60 \skewchar\sixsy='60
\hyphenchar\ninett=-1 \hyphenchar\eighttt=-1 \hyphenchar\tentt=-1
\def\bfmath{\cmmib}                 
\font\tencmmib=cmmib10  \newfam\cmmibfam  \skewchar\tencmmib='177
\font\tencmbsy=cmbsy10  \newfam\cmbsyfam  \skewchar\tencmbsy='60
\def\scaps{\cmcsc}                 
\font\tencmcsc=cmcsc10  \newfam\cmcscfam
\ifnum\ingrandimento=1095

\font\capsone=cmcsc10 at 10.95pt 

\else

\font\capsone=cmcsc10 at 12pt 
\fi

\def\ttaarr{\bf}		
\def\ppaarr{\sl}		

%
%
%
\newfam\eufmfam \newfam\msamfam \newfam\msbmfam \newfam\eufbfam
\def\Loadeulerfonts{\global\contaeuler=1 \ifx\arisposta\amsrisposta
\font\teneufm=eufm10              
\font\eighteufm=eufm8 \font\nineeufm=eufm9 \font\sixeufm=eufm6
\font\seveneufm=eufm7  \font\fiveeufm=eufm5
\font\teneufb=eufb10              
\font\eighteufb=eufb8 \font\nineeufb=eufb9 \font\sixeufb=eufb6
\font\seveneufb=eufb7  \font\fiveeufb=eufb5
\font\teneurm=eurm10              
\font\eighteurm=eurm8 \font\nineeurm=eurm9
\font\teneurb=eurb10              
\font\eighteurb=eurb8 \font\nineeurb=eurb9
\font\teneusm=eusm10              
\font\eighteusm=eusm8 \font\nineeusm=eusm9
\font\teneusb=eusb10              
\font\eighteusb=eusb8 \font\nineeusb=eusb9
\else \def\eufm{\tt} \def\eufb{\tt} \def\eurm{\tt} \def\eurb{\tt}
\def\eusm{\tt} \def\eusb{\tt}    \fi}
\def\loadeuler{\Loadeulerfonts\tenpoint}
\def\loadamsmath{\global\contaams=1 \ifx\arisposta\amsrisposta
\font\tenmsam=msam10 \font\ninemsam=msam9 \font\eightmsam=msam8
\font\sevenmsam=msam7 \font\sixmsam=msam6 \font\fivemsam=msam5
\font\tenmsbm=msbm10 \font\ninemsbm=msbm9 \font\eightmsbm=msbm8
\font\sevenmsbm=msbm7 \font\sixmsbm=msbm6 \font\fivemsbm=msbm5
\else \def\msbm{\bf} \fi \def\Bbb{\msbm} \def\symbl{\msam} \tenpoint}
\def\loadcyrill{\global\contacyrill=1 \ifx\arisposta\amsrisposta
\font\tenwncyr=wncyr10 \font\ninewncyr=wncyr9 \font\eightwncyr=wncyr8
\font\tenwncyb=wncyr10 \font\ninewncyb=wncyr9 \font\eightwncyb=wncyr8
\font\tenwncyi=wncyr10 \font\ninewncyi=wncyr9 \font\eightwncyi=wncyr8
\else \def\cyrill{\sl} \def\cyrilb{\sl} \def\cyrili{\sl} \fi\tenpoint}
\ifx\arisposta\amsrisposta
\font\sevenex=cmex7               
\font\eightex=cmex8  \font\nineex=cmex9
\font\ninecmmib=cmmib9   \font\eightcmmib=cmmib8
\font\sevencmmib=cmmib7 \font\sixcmmib=cmmib6
\font\fivecmmib=cmmib5   \skewchar\ninecmmib='177
\skewchar\eightcmmib='177  \skewchar\sevencmmib='177
\skewchar\sixcmmib='177   \skewchar\fivecmmib='177
\font\ninecmbsy=cmbsy9    \font\eightcmbsy=cmbsy8
\font\sevencmbsy=cmbsy7  \font\sixcmbsy=cmbsy6
\font\fivecmbsy=cmbsy5   \skewchar\ninecmbsy='60
\skewchar\eightcmbsy='60  \skewchar\sevencmbsy='60
\skewchar\sixcmbsy='60    \skewchar\fivecmbsy='60
\font\ninecmcsc=cmcsc9    \font\eightcmcsc=cmcsc8     \else
\def\cmmib{\fam\cmmibfam\tencmmib}\textfont\cmmibfam=\tencmmib
\scriptfont\cmmibfam=\tencmmib \scriptscriptfont\cmmibfam=\tencmmib
\def\cmbsy{\fam\cmbsyfam\tencmbsy} \textfont\cmbsyfam=\tencmbsy
\scriptfont\cmbsyfam=\tencmbsy \scriptscriptfont\cmbsyfam=\tencmbsy
\scriptfont\cmcscfam=\tencmcsc \scriptscriptfont\cmcscfam=\tencmcsc
\def\cmcsc{\fam\cmcscfam\tencmcsc} \textfont\cmcscfam=\tencmcsc \fi
\catcode`@=11
\newskip\ttglue
\gdef\tenpoint{\def\rm{\fam0\tenrm}
  \textfont0=\tenrm \scriptfont0=\sevenrm \scriptscriptfont0=\fiverm
  \textfont1=\teni \scriptfont1=\seveni \scriptscriptfont1=\fivei
  \textfont2=\tensy \scriptfont2=\sevensy \scriptscriptfont2=\fivesy
  \textfont3=\tenex \scriptfont3=\tenex \scriptscriptfont3=\tenex
  \def\mcal{\fam2 \tensy}  \def\mmit{\fam1 \teni}
  \textfont\itfam=\tenit \def\it{\fam\itfam\tenit}
  \textfont\slfam=\tensl \def\sl{\fam\slfam\tensl}
  \textfont\ttfam=\tentt \scriptfont\ttfam=\eighttt
  \scriptscriptfont\ttfam=\eighttt  \def\tt{\fam\ttfam\tentt}
  \textfont\bffam=\tenbf \scriptfont\bffam=\sevenbf
  \scriptscriptfont\bffam=\fivebf \def\bf{\fam\bffam\tenbf}
     \ifx\arisposta\amsrisposta    \ifnum\contaeuler=1
  \textfont\eufmfam=\teneufm \scriptfont\eufmfam=\seveneufm
  \scriptscriptfont\eufmfam=\fiveeufm \def\eufm{\fam\eufmfam\teneufm}
  \textfont\eufbfam=\teneufb \scriptfont\eufbfam=\seveneufb
  \scriptscriptfont\eufbfam=\fiveeufb \def\eufb{\fam\eufbfam\teneufb}
  \def\eurm{\teneurm} \def\eurb{\teneurb} \def\eusm{\teneusm}
  \def\eusb{\teneusb}    \fi    \ifnum\contaams=1
  \textfont\msamfam=\tenmsam \scriptfont\msamfam=\sevenmsam
  \scriptscriptfont\msamfam=\fivemsam \def\msam{\fam\msamfam\tenmsam}
  \textfont\msbmfam=\tenmsbm \scriptfont\msbmfam=\sevenmsbm
  \scriptscriptfont\msbmfam=\fivemsbm \def\msbm{\fam\msbmfam\tenmsbm}
     \fi      \ifnum\contacyrill=1     \def\cyrill{\tenwncyr}
  \def\cyrilb{\tenwncyb}  \def\cyrili{\tenwncyi}         \fi
  \textfont3=\tenex \scriptfont3=\sevenex \scriptscriptfont3=\sevenex
  \def\cmmib{\fam\cmmibfam\tencmmib} \scriptfont\cmmibfam=\sevencmmib
  \textfont\cmmibfam=\tencmmib  \scriptscriptfont\cmmibfam=\fivecmmib
  \def\cmbsy{\fam\cmbsyfam\tencmbsy} \scriptfont\cmbsyfam=\sevencmbsy
  \textfont\cmbsyfam=\tencmbsy  \scriptscriptfont\cmbsyfam=\fivecmbsy
  \def\cmcsc{\fam\cmcscfam\tencmcsc} \scriptfont\cmcscfam=\eightcmcsc
  \textfont\cmcscfam=\tencmcsc \scriptscriptfont\cmcscfam=\eightcmcsc
     \fi            \tt \ttglue=.5em plus.25em minus.15em
  \normalbaselineskip=12pt
  \setbox\strutbox=\hbox{\vrule height8.5pt depth3.5pt width0pt}
  \let\sc=\eightrm \let\big=\tenbig   \normalbaselines
  \baselineskip=\infralinea  \rm}
\gdef\ninepoint{\def\rm{\fam0\ninerm}
  \textfont0=\ninerm \scriptfont0=\sixrm \scriptscriptfont0=\fiverm
  \textfont1=\ninei \scriptfont1=\sixi \scriptscriptfont1=\fivei
  \textfont2=\ninesy \scriptfont2=\sixsy \scriptscriptfont2=\fivesy
  \textfont3=\tenex \scriptfont3=\tenex \scriptscriptfont3=\tenex
  \def\mcal{\fam2 \ninesy}  \def\mmit{\fam1 \ninei}
  \textfont\itfam=\nineit \def\it{\fam\itfam\nineit}
  \textfont\slfam=\ninesl \def\sl{\fam\slfam\ninesl}
  \textfont\ttfam=\ninett \scriptfont\ttfam=\eighttt
  \scriptscriptfont\ttfam=\eighttt \def\tt{\fam\ttfam\ninett}
  \textfont\bffam=\ninebf \scriptfont\bffam=\sixbf
  \scriptscriptfont\bffam=\fivebf \def\bf{\fam\bffam\ninebf}
     \ifx\arisposta\amsrisposta  \ifnum\contaeuler=1
  \textfont\eufmfam=\nineeufm \scriptfont\eufmfam=\sixeufm
  \scriptscriptfont\eufmfam=\fiveeufm \def\eufm{\fam\eufmfam\nineeufm}
  \textfont\eufbfam=\nineeufb \scriptfont\eufbfam=\sixeufb
  \scriptscriptfont\eufbfam=\fiveeufb \def\eufb{\fam\eufbfam\nineeufb}
  \def\eurm{\nineeurm} \def\eurb{\nineeurb} \def\eusm{\nineeusm}
  \def\eusb{\nineeusb}     \fi   \ifnum\contaams=1
  \textfont\msamfam=\ninemsam \scriptfont\msamfam=\sixmsam
  \scriptscriptfont\msamfam=\fivemsam \def\msam{\fam\msamfam\ninemsam}
  \textfont\msbmfam=\ninemsbm \scriptfont\msbmfam=\sixmsbm
  \scriptscriptfont\msbmfam=\fivemsbm \def\msbm{\fam\msbmfam\ninemsbm}
     \fi       \ifnum\contacyrill=1     \def\cyrill{\ninewncyr}
  \def\cyrilb{\ninewncyb}  \def\cyrili{\ninewncyi}         \fi
  \textfont3=\nineex \scriptfont3=\sevenex \scriptscriptfont3=\sevenex
  \def\cmmib{\fam\cmmibfam\ninecmmib}  \textfont\cmmibfam=\ninecmmib
  \scriptfont\cmmibfam=\sixcmmib \scriptscriptfont\cmmibfam=\fivecmmib
  \def\cmbsy{\fam\cmbsyfam\ninecmbsy}  \textfont\cmbsyfam=\ninecmbsy
  \scriptfont\cmbsyfam=\sixcmbsy \scriptscriptfont\cmbsyfam=\fivecmbsy
  \def\cmcsc{\fam\cmcscfam\ninecmcsc} \scriptfont\cmcscfam=\eightcmcsc
  \textfont\cmcscfam=\ninecmcsc \scriptscriptfont\cmcscfam=\eightcmcsc
     \fi            \tt \ttglue=.5em plus.25em minus.15em
  \normalbaselineskip=11pt
  \setbox\strutbox=\hbox{\vrule height8pt depth3pt width0pt}
  \let\sc=\sevenrm \let\big=\ninebig \normalbaselines\rm}
\gdef\eightpoint{\def\rm{\fam0\eightrm}
  \textfont0=\eightrm \scriptfont0=\sixrm \scriptscriptfont0=\fiverm
  \textfont1=\eighti \scriptfont1=\sixi \scriptscriptfont1=\fivei
  \textfont2=\eightsy \scriptfont2=\sixsy \scriptscriptfont2=\fivesy
  \textfont3=\tenex \scriptfont3=\tenex \scriptscriptfont3=\tenex
  \def\mcal{\fam2 \eightsy}  \def\mmit{\fam1 \eighti}
  \textfont\itfam=\eightit \def\it{\fam\itfam\eightit}
  \textfont\slfam=\eightsl \def\sl{\fam\slfam\eightsl}
  \textfont\ttfam=\eighttt \scriptfont\ttfam=\eighttt
  \scriptscriptfont\ttfam=\eighttt \def\tt{\fam\ttfam\eighttt}
  \textfont\bffam=\eightbf \scriptfont\bffam=\sixbf
  \scriptscriptfont\bffam=\fivebf \def\bf{\fam\bffam\eightbf}
     \ifx\arisposta\amsrisposta   \ifnum\contaeuler=1
  \textfont\eufmfam=\eighteufm \scriptfont\eufmfam=\sixeufm
  \scriptscriptfont\eufmfam=\fiveeufm \def\eufm{\fam\eufmfam\eighteufm}
  \textfont\eufbfam=\eighteufb \scriptfont\eufbfam=\sixeufb
  \scriptscriptfont\eufbfam=\fiveeufb \def\eufb{\fam\eufbfam\eighteufb}
  \def\eurm{\eighteurm} \def\eurb{\eighteurb} \def\eusm{\eighteusm}
  \def\eusb{\eighteusb}       \fi    \ifnum\contaams=1
  \textfont\msamfam=\eightmsam \scriptfont\msamfam=\sixmsam
  \scriptscriptfont\msamfam=\fivemsam \def\msam{\fam\msamfam\eightmsam}
  \textfont\msbmfam=\eightmsbm \scriptfont\msbmfam=\sixmsbm
  \scriptscriptfont\msbmfam=\fivemsbm \def\msbm{\fam\msbmfam\eightmsbm}
     \fi       \ifnum\contacyrill=1     \def\cyrill{\eightwncyr}
  \def\cyrilb{\eightwncyb}  \def\cyrili{\eightwncyi}         \fi
  \textfont3=\eightex \scriptfont3=\sevenex \scriptscriptfont3=\sevenex
  \def\cmmib{\fam\cmmibfam\eightcmmib}  \textfont\cmmibfam=\eightcmmib
  \scriptfont\cmmibfam=\sixcmmib \scriptscriptfont\cmmibfam=\fivecmmib
  \def\cmbsy{\fam\cmbsyfam\eightcmbsy}  \textfont\cmbsyfam=\eightcmbsy
  \scriptfont\cmbsyfam=\sixcmbsy \scriptscriptfont\cmbsyfam=\fivecmbsy
  \def\cmcsc{\fam\cmcscfam\eightcmcsc} \scriptfont\cmcscfam=\eightcmcsc
  \textfont\cmcscfam=\eightcmcsc \scriptscriptfont\cmcscfam=\eightcmcsc
     \fi             \tt \ttglue=.5em plus.25em minus.15em
  \normalbaselineskip=9pt
  \setbox\strutbox=\hbox{\vrule height7pt depth2pt width0pt}
  \let\sc=\sixrm \let\big=\eightbig \normalbaselines\rm }
\gdef\tenbig#1{{\hbox{$\left#1\vbox to8.5pt{}\right.\n@space$}}}
\gdef\ninebig#1{{\hbox{$\textfont0=\tenrm\textfont2=\tensy
   \left#1\vbox to7.25pt{}\right.\n@space$}}}
\gdef\eightbig#1{{\hbox{$\textfont0=\ninerm\textfont2=\ninesy
   \left#1\vbox to6.5pt{}\right.\n@space$}}}
\def\alternativefont#1#2{\ifx\arisposta\amsrisposta \relax \else
\xdef#1{#2} \fi}
\global\contaeuler=0 \global\contacyrill=0 \global\contaams=0
%
%
%
%
\newbox\fotlinebb \newbox\hedlinebb \newbox\leftcolumn
\gdef\makeheadline{\vbox to 0pt{\vskip-22.5pt
     \fullline{\vbox to8.5pt{}\the\headline}\vss}\nointerlineskip}
\gdef\makehedlinebb{\vbox to 0pt{\vskip-22.5pt
     \fullline{\vbox to8.5pt{}\copy\hedlinebb\hfil
     \line{\hfill\the\headline\hfill}}\vss} \nointerlineskip}
\gdef\makefootline{\baselineskip=24pt \fullline{\the\footline}}
\gdef\makefotlinebb{\baselineskip=24pt
    \fullline{\copy\fotlinebb\hfil\line{\hfill\the\footline\hfill}}}
\gdef\doubleformat{\shipout\vbox{\Landspec\makehedlinebb
     \fullline{\box\leftcolumn\hfil\columnbox}\makefotlinebb}
     \advancepageno}
\gdef\columnbox{\leftline{\pagebody}}
\gdef\line#1{\hbox to\hsize{\hskip\leftskip#1\hskip\rightskip}}
\gdef\fullline#1{\hbox to\fullhsize{\hskip\leftskip{#1}%
\hskip\rightskip}}
\gdef\footnote#1{\let\@sf=\empty
         \ifhmode\edef\#sf{\spacefactor=\the\spacefactor}\/\fi
         #1\@sf\vfootnote{#1}}
\gdef\vfootnote#1{\insert\footins\bgroup
         \ifnum\dimnota=1  \eightpoint\fi
         \ifnum\dimnota=2  \ninepoint\fi
         \ifnum\dimnota=0  \tenpoint\fi
         \interlinepenalty=\interfootnotelinepenalty
         \splittopskip=\ht\strutbox
         \splitmaxdepth=\dp\strutbox \floatingpenalty=20000
         \leftskip=\oldssposta \rightskip=\olddsposta
         \spaceskip=0pt \xspaceskip=0pt
         \ifnum\sinnota=0   \textindent{#1}\fi
         \ifnum\sinnota=1   \item{#1}\fi
         \footstrut\futurelet\next\fo@t}
\gdef\fo@t{\ifcat\bgroup\noexpand\next \let\next\f@@t
             \else\let\next\f@t\fi \next}
\gdef\f@@t{\bgroup\aftergroup\@foot\let\next}
\gdef\f@t#1{#1\@foot} \gdef\@foot{\strut\egroup}
\gdef\footstrut{\vbox to\splittopskip{}}
\skip\footins=\bigskipamount
\count\footins=1000  \dimen\footins=8in
\catcode`@=12
\tenpoint
\ifnum\unoduecol=1 \hsize=\tothsize   \fullhsize=\tothsize \fi
\ifnum\unoduecol=2 \hsize=\collhsize  \fullhsize=\tothsize \fi
\global\let\lrcol=L      \ifnum\unoduecol=1
\output{\plainoutput{\ifnum\tipbnota=2 \clearnmbnota\fi}} \fi
\ifnum\unoduecol=2 \output{\if L\lrcol
     \global\setbox\leftcolumn=\columnbox
     \global\setbox\fotlinebb=\line{\hfill\the\footline\hfill}
     \global\setbox\hedlinebb=\line{\hfill\the\headline\hfill}
     \advancepageno  \global\let\lrcol=R
     \else  \doubleformat \global\let\lrcol=L \fi
     \ifnum\outputpenalty>-20000 \else\dosupereject\fi
     \ifnum\tipbnota=2\clearnmbnota\fi }\fi
\def\ifdoublepage{\ifnum\unoduecol=2 }
\gdef\yespagenumbers{\footline={\hss\tenrm\folio\hss}}
\gdef\ciao{ \ifnum\fdefcontre=1 \endfdef\fi
     \par\vfill\supereject \ifnum\unoduecol=2
     \if R\lrcol  \headline={}\nopagenumbers\null\vfill\eject
     \fi\fi \end}

\newskip\olddsposta \newskip\oldssposta
\global\oldssposta=\leftskip \global\olddsposta=\rightskip

\def\filldots{\leaders\hbox to 1em{\hss.\hss}\hfill}
\def\inquadrb#1 {\vbox {\hrule  \hbox{\vrule \vbox {\vskip .2cm
    \hbox {\ #1\ } \vskip .2cm } \vrule  }  \hrule} }
 \def\newline{\hfil\break}
\def\jump{\vskip\baselineskip} \newskip\iinnffrr
\def\sjump{\iinnffrr=\baselineskip
          \divide\iinnffrr by 2 \vskip\iinnffrr}
\def\bjump{\vskip\baselineskip \vskip\baselineskip}
\newcount\nmbnota  \def\clearnmbnota{\global\nmbnota=0}
\newcount\tipbnota \def\letterfootnote{\global\tipbnota=1}

\def\note#1{\global\advance\nmbnota by 1 \ifnum\tipbnota=1
    \footnote{$^{\rm\nttlett}$}{#1} \else {\ifnum\tipbnota=2
    \footnote{$^{\nttsymb}$}{#1}
    \else\footnote{$^{\the\nmbnota}$}{#1}\fi}\fi}
\def\nttlett{\ifcase\nmbnota \or a\or b\or c\or d\or e\or f\or
g\or h\or i\or j\or k\or l\or m\or n\or o\or p\or q\or r\or
s\or t\or u\or v\or w\or y\or x\or z\fi}
\def\nttsymb{\ifcase\nmbnota \or\dag\or\sharp\or\ddag\or\star\or
\natural\or\flat\or\clubsuit\or\diamondsuit\or\heartsuit
\or\spadesuit\fi}   \clearnmbnota
\def\numberfootnote{\global\tipbnota=0} \numberfootnote
\def\setnote#1{\expandafter\xdef\csname#1\endcsname{
\ifnum\tipbnota=1 {\rm\nttlett} \else {\ifnum\tipbnota=2
{\nttsymb} \else \the\nmbnota\fi}\fi} }
\newcount\nbmfig  \def\clearnbmfig{\global\nbmfig=0}
\gdef\figure{\global\advance\nbmfig by 1
      {\rm fig. \the\nbmfig}}   \clearnbmfig
\def\setfig#1{\expandafter\xdef\csname#1\endcsname{fig. \the\nbmfig}}
 \def\endformula{\eqno\numero $$}
 \def\efr{\endformula}
\newcount\frmcount \def\clearfrmcount{\global\frmcount=0}
\def\numero{\global\advance\frmcount by 1   \ifnum\indappcount=0
  {\ifnum\cpcount <1 {\hbox{\rm (\the\frmcount )}}  \else
  {\hbox{\rm (\the\cpcount .\the\frmcount )}} \fi}  \else
  {\hbox{\rm (\applett .\the\frmcount )}} \fi}
\def\nameformula#1{\global\advance\frmcount by 1%
\ifnum\draftnum=0  {\ifnum\indappcount=0%
{\ifnum\cpcount<1\xdef\spzzttrra{(\the\frmcount )}%
\else\xdef\spzzttrra{(\the\cpcount .\the\frmcount )}\fi}%
\else\xdef\spzzttrra{(\applett .\the\frmcount )}\fi}%
\else\xdef\spzzttrra{(#1)}\fi%
\expandafter\xdef\csname#1\endcsname{\spzzttrra}
\eqno \hbox{\rm\spzzttrra} $$}
\def\nfr{\nameformula}    
\def\nameali#1{\global\advance\frmcount by 1%
\ifnum\draftnum=0  {\ifnum\indappcount=0%
{\ifnum\cpcount<1\xdef\spzzttrra{(\the\frmcount )}%
\else\xdef\spzzttrra{(\the\cpcount .\the\frmcount )}\fi}%
\else\xdef\spzzttrra{(\applett .\the\frmcount )}\fi}%
\else\xdef\spzzttrra{(#1)}\fi%
\expandafter\xdef\csname#1\endcsname{\spzzttrra}
  \hbox{\rm\spzzttrra} }      \clearfrmcount
\newcount\cpcount \def\clearcpcount{\global\cpcount=0}
\newcount\subcpcount \def\clearsubcpcount{\global\subcpcount=0}
\newcount\appcount \def\clearappcount{\global\appcount=0}
\newcount\indappcount \def\clearindappcount{\indappcount=0}
\newcount\sottoparcount 

\def\applett{\ifcase\appcount  \or {A}\or {B}\or {C}\or
{D}\or {E}\or {F}\or {G}\or {H}\or {I}\or {J}\or {K}\or {L}\or
{M}\or {N}\or {O}\or {P}\or {Q}\or {R}\or {S}\or {T}\or {U}\or
{V}\or {W}\or {X}\or {Y}\or {Z}\fi    \ifnum\appcount<0
\immediate\write16 {Panda ERROR - Appendix: counter "appcount"
out of range}\fi  \ifnum\appcount>26  \immediate\write16 {Panda
ERROR - Appendix: counter "appcount" out of range}\fi}
\clearappcount  \clearindappcount \newcount\connttrre
\def\clearconnttrre{\global\connttrre=0} \newcount\countref
\def\clearcountref{\global\countref=0} \clearcountref
\def\chapter#1{\global\advance\cpcount by 1 \clearfrmcount
                 \goodbreak\null\vbox{\jump\nobreak
                 \clearsubcpcount\clearindappcount
                 \itemitem{\ttaarr\the\cpcount .\qquad}{\ttaarr #1}
                 \par\nobreak\jump\sjump}\nobreak}
\def\section#1{\global\advance\subcpcount by 1 \goodbreak\null
               \vbox{\sjump\nobreak\ifnum\indappcount=0
                 {\ifnum\cpcount=0 {\itemitem{\ppaarr
               .\the\subcpcount\quad\enskip\ }{\ppaarr #1}\par} \else
                 {\itemitem{\ppaarr\the\cpcount .\the\subcpcount\quad
                  \enskip\ }{\ppaarr #1} \par}  \fi}
                \else{\itemitem{\ppaarr\applett .\the\subcpcount\quad
                 \enskip\ }{\ppaarr #1}\par}\fi\nobreak\jump}\nobreak}
\clearsubcpcount
\def\appendix#1{\global\advance\appcount by 1 \clearfrmcount
                  \goodbreak\null\vbox{\jump\nobreak
                  \global\advance\indappcount by 1 \clearsubcpcount
          \itemitem{ }{\hskip-40pt\ttaarr Appendix\ #1}
             \nobreak\jump\sjump}\nobreak}
\clearappcount \clearindappcount
\def\references{\goodbreak\null\vbox{\jump\nobreak
   \itemitem{}{\ttaarr References} \nobreak\jump\sjump}\nobreak}

\clearcpcount\clearcountref

\def\setchap#1{\ifnum\indappcount=0{\ifnum\subcpcount=0%
\xdef\spzzttrra{\the\cpcount}%
\else\xdef\spzzttrra{\the\cpcount .\the\subcpcount}\fi}
\else{\ifnum\subcpcount=0 \xdef\spzzttrra{\applett}%
\else\xdef\spzzttrra{\applett .\the\subcpcount}\fi}\fi
\expandafter\xdef\csname#1\endcsname{\spzzttrra}}
\newcount\draftnum \newcount\ppora   \newcount\ppminuti
\global\ppora=\time   \global\ppminuti=\time
\global\divide\ppora by 60  \draftnum=\ppora
\multiply\draftnum by 60    \global\advance\ppminuti by -\draftnum
\def\droggi{\number\day /\number\month /\number\year\ \the\ppora
:\the\ppminuti}     \global\draftnum=0
\def\draftcomment#1{\ifnum\draftnum=0 \relax \else
{\ {\bf ***}\ #1\ {\bf ***}\ }\fi} 
%
%
\catcode`@=11
\gdef\Ref#1{\expandafter\ifx\csname @rrxx@#1\endcsname\relax%
{\global\advance\countref by 1    \ifnum\countref>200
\immediate\write16 {Panda ERROR - Ref: maximum number of references
exceeded}  \expandafter\xdef\csname @rrxx@#1\endcsname{0}\else
\expandafter\xdef\csname @rrxx@#1\endcsname{\the\countref}\fi}\fi
\ifnum\draftnum=0 \csname @rrxx@#1\endcsname \else#1\fi}
\gdef\beginref{\ifnum\draftnum=0  \gdef\Rref{\fairef}
\gdef\endref{\scriviref} \else\relax\fi
\ifx\risposta\mplarisposta \ninepoint \fi
\parskip 2pt plus.2pt \baselineskip=12pt}
\def\Reflab#1{[#1]} \gdef\Rref#1#2{\item{\Reflab{#1}}{#2}}
\gdef\endref{\relax}  \newcount\conttemp
\gdef\fairef#1#2{\expandafter\ifx\csname @rrxx@#1\endcsname\relax
{\global\conttemp=0 \immediate\write16 {Panda ERROR - Ref: reference
[#1] undefined}} \else
{\global\conttemp=\csname @rrxx@#1\endcsname } \fi
\global\advance\conttemp by 50  \global\setbox\conttemp=\hbox{#2} }
\gdef\scriviref{\clearconnttrre\conttemp=50
\loop\ifnum\connttrre<\countref \advance\conttemp by 1
\advance\connttrre by 1
\item{\Reflab{\the\connttrre}}{\unhcopy\conttemp} \repeat}
\clearcountref \clearconnttrre
\catcode`@=12
\ifx\risposta\mplarisposta \def\Reflab#1{#1.} \letterfootnote \fi

\def\slashchar#1{\setbox0=\hbox{$#1$} \dimen0=\wd0
     \setbox1=\hbox{/} \dimen1=\wd1 \ifdim\dimen0>\dimen1
      \rlap{\hbox to \dimen0{\hfil/\hfil}} #1 \else
      \rlap{\hbox to \dimen1{\hfil$#1$\hfil}} / \fi}
\ifx\oldchi\undefined \let\oldchi=\chi
  \def\cchi{{\raise 1pt\hbox{$\oldchi$}}} \let\chi=\cchi \fi
  
\def\del{\partial}   

\def\frac#1#2{{\textstyle{#1 \over #2}}}

\def\half{\ifinner {\scriptstyle {1 \over 2}}\else {1 \over 2} \fi}

\def\simge{\rlap{\raise 2pt \hbox{$>$}}{\lower 2pt \hbox{$\sim$}}}
\def\simle{\rlap{\raise 2pt \hbox{$<$}}{\lower 2pt \hbox{$\sim$}}}

\def\vbig#1#2{{\vbigd@men=#2\divide\vbigd@men by 2%
\hbox{$\left#1\vbox to \vbigd@men{}\right.\n@space$}}}

%
%
\newcount\fdefcontre \newcount\fdefcount \newcount\indcount
\newread\filefdef  \newread\fileftmp  \newwrite\filefdef
\newwrite\fileftmp     \def\strip#1*.A {#1}
\def\futuredef#1{\beginfdef
\expandafter\ifx\csname#1\endcsname\relax%
{\immediate\write\fileftmp {#1*.A}
\immediate\write16 {Panda Warning - fdef: macro "#1" on page
\the\pageno \space undefined}
\ifnum\draftnum=0 \expandafter\xdef\csname#1\endcsname{(?)}
\else \expandafter\xdef\csname#1\endcsname{(#1)} \fi
\global\advance\fdefcount by 1}\fi   \csname#1\endcsname}

\def\beginfdef{\ifnum\fdefcontre=0
\immediate\openin\filefdef \jobname.fdef
\immediate\openout\fileftmp \jobname.ftmp
\global\fdefcontre=1  \ifeof\filefdef \immediate\write16 {Panda
WARNING - fdef: file \jobname.fdef not found, run TeX again}
\else \immediate\read\filefdef to\spzzttrra
\global\advance\fdefcount by \spzzttrra
\indcount=0      \loop\ifnum\indcount<\fdefcount
\advance\indcount by 1   \immediate\read\filefdef to\spezttrra
\immediate\read\filefdef to\sppzttrra
\edef\spzzttrra{\expandafter\strip\spezttrra}
\immediate\write\fileftmp {\spzzttrra *.A}
\expandafter\xdef\csname\spzzttrra\endcsname{\sppzttrra}
\repeat \fi \immediate\closein\filefdef \fi}
\def\endfdef{\immediate\closeout\fileftmp   \ifnum\fdefcount>0
\immediate\openin\fileftmp \jobname.ftmp
\immediate\openout\filefdef \jobname.fdef
\immediate\write\filefdef {\the\fdefcount}   \indcount=0
\loop\ifnum\indcount<\fdefcount    \advance\indcount by 1
\immediate\read\fileftmp to\spezttrra
\edef\spzzttrra{\expandafter\strip\spezttrra}
\immediate\write\filefdef{\spzzttrra *.A}
\edef\spezttrra{\string{\csname\spzzttrra\endcsname\string}}
\iwritel\filefdef{\spezttrra}
\repeat  \immediate\closein\fileftmp \immediate\closeout\filefdef
\immediate\write16 {Panda Warning - fdef: Label(s) may have changed,
re-run TeX to get them right}\fi}
\def\iwritel#1#2{\newlinechar=-1
{\newlinechar=`\ \immediate\write#1{#2}}\newlinechar=-1}
\global\fdefcontre=0 \global\fdefcount=0 \global\indcount=0
%
%
\null
%
%
%
%

%
\loadamsmath
\loadeuler
\mathchardef\bbalpha="710B
\mathchardef\bbbeta="710C
\mathchardef\bbgamma="710D
\mathchardef\bbomega="7121
\mathchardef\bbmu="7216
\mathchardef\bbnu="7217
\mathchardef\bbrho="721A
\mathchardef\bbxi="7218
\mathchardef\bbta="7211
\def\balpha{{\bfmath\bbalpha}}

\def\bomega{{\bfmath\bbomega}}
\def\bmu{{\bfmath\bbmu}}
\def\bnu{{\bfmath\bbnu}}
\def\brho{{\bfmath\bbrho}}
\def\bxi{{\bfmath\bbxi}}
\def\bta{{\bfmath\bbta}}

\def\bq{{\bfmath q}}
\def\bH{{\bfmath H}}
\def\sunlr{ {\rm SU} (n)_{\rm L} \times {\rm SU} (n)_{\rm R} }
\def\glr{ G_{\rm L} \times G_{\rm R} }
\pageno=0\baselineskip=14pt
\nopagenumbers{
\line{\hfill CERN-TH/95-271}
\line{\hfill DAMTP/96-33}
\line{\hfill SWAT/103}
\line{\hfill\tt hep-th/9603190}
\ifdoublepage \bjump\bjump\else\vfill\fi
\centerline{\capsone Exact scattering in the SU($n$) supersymmetric}
\sjump\sjump
\centerline{\capsone principal chiral model}
\bjump
\centerline{\scaps Jonathan M. Evans$^{*}$\note{E-mail: 
{\tt j.m.evans@damtp.cam.ac.uk} }
and Timothy J. Hollowood$^\ddagger$\note{E-mail: 
{\tt t.hollowood@swansea.ac.uk} }
}
\sjump
\sjump
\centerline{\sl $^*$Theory Division, CERN, CH-1211 Geneva 23, Switzerland}
\sjump
\centerline{\sl $^*$DAMTP, University of Cambridge, Silver Street, 
Cambridge CB3 9EW, U.K.\note{Present address} }
\sjump
\centerline{\sl $^\ddagger$Department of Physics, University of Wales Swansea,}
\centerline{\sl Singleton Park, Swansea SA2 8PP, U.K.}
\bjump
\ifdoublepage
\vfill
\noindent
\line{CERN-TH/95-271}
\line{March 1996\hfill}
\eject\null\vfill\fi
\centerline{\capsone ABSTRACT}\sjump

The complete spectrum of states in the
supersymmetric principal chiral model based on SU$(n)$ is conjectured,
and an exact factorizable S-matrix is proposed to describe
scattering amongst these states. The ${\rm SU}(n)_{\rm L}
\times{\rm SU}(n)_{\rm R}$ symmetry of the lagrangian is manifest in
the S-matrix construction. The supersymmetries, on the other hand,
are incorporated in the guise of spin-1/2 charges acting on a set of
RSOS kinks associated with $su(n)$ at level $n$. 
To test the proposed S-matrix, calculations of the change in the ground-state
energy in the presence of a coupling to a background 
charge are carried out. The results derived from the lagrangian using 
perturbation theory and from the S-matrix using the TBA 
are found to be in
complete agreement for a variety of background charges which pick out,
in turn, the highest weight states in each of the
fundamental representations of SU$(n)$. 
In particular, these methods rule out the possibility of additional 
CDD factors  
in the S-matrix. Comparison of the expressions found for the
free-energy also yields an exact result for the
mass-gap in these models:
$m/\Lambda_{\overline{\rm MS}} = (n / \pi ) \sin ( \pi / n)$.

\sjump\vfill
\ifdoublepage \else
\noindent
\line{March 1996\hfill}\fi
\eject}
\yespagenumbers\pageno=1
%
%

\chapter{Introduction}

Sigma-models in two dimensions have played an important r\^ole in 
helping us improve our understanding of quantum field theory;
in particular, they have provided
illuminating examples of integrable yet highly non-trivial 
quantum systems in which one can investigate non-perturbative
phenomena in an explicit way. 
It seems that a sigma-model based on a Riemannian symmetric
space $G/H$ (with $G$ and $H$ compact Lie groups) 
is always classically integrable, but that 
integrability may be spoiled by anomalies 
at the quantum level unless $H$ is simple (see [\Ref{AAR}] and
references given there). 
A prominent subset of the class of quantum-integrable sigma-models consists
of the principal chiral
models (PCMs). The target space for a PCM is
some simple, compact Lie group, $G$---which can of course be regarded 
as a symmetric space $G\times G/G$---so the basic 
field in the theory is a 
$G$-valued function $U(x,t)$ on Minkowski space. 
The lagrangian governing its behaviour is 
$$
{\cal L}={1\over g}{\rm Tr}\left(\partial_\mu U^{-1}\partial^\mu U\right) ,
\nfr{BLAG}
where $g$ is a dimensionless coupling, and the theory 
is clearly invariant under a global symmetry group $\glr$ 
which acts by left and right multiplication:
$ U \rightarrow g^{\phantom{1}}_{\rm L} U g^{-1}_{\rm R} $ for 
any $g_{\rm L}$ and $g_{\rm R}$ in $G$.

In four dimensions, lagrangians such as \BLAG\ have been studied for 
many years:
they are non-renormalizable but they are nevertheless useful 
in phenomenological, low-energy
descriptions of strong interactions in which the chiral symmetry
$\glr$ is spontaneously
broken to a diagonal flavour subgroup $G$.
In two dimensions---which is the case of interest here---such
lagrangians are, by contrast, 
renormalizable, and spontaneous symmetry breaking is not allowed
on general grounds [\Ref{Col}].
Furthermore, since these two-dimensional models are integrable in the
sense that they possess 
infinitely many conserved charges, their S-matrices
must factorize, and
by enforcing this powerful constraint 
in conjunction with the usual axioms of S-matrix theory, along
with the existence of the unbroken $\glr$ symmetry, 
one can hope to determine the
exact expressions for all scattering amplitudes.

For the cases where $G$ is a classical Lie group, such S-matrices were 
conjectured some time ago [\Ref{ORW},\Ref{AAL}].
The spectrum
is postulated to consist of particles transforming in representations
$(R,\bar R)$ of $\glr$, 
where $R$ is some (possibly reducible) representation of $G$.
In the case $G$ = SU($n$) there are a total of $(n{-}1)$ multiplets of
particles, each associated with the fundamental representations
$R = R_a$ of $G$ with $a = 1, \ldots , n{-}1$ and 
with corresponding masses given by
$$
m_a=m{\sin(\pi a/n)\over\sin(\pi/n)},\qquad a=1,2,\ldots,n-1,
\nfr{MASS} 
where $m$ is the mass of the lightest state.\note{We use the obvious 
labelling for the fundamental
representations, agreeing with [\Ref{AAL}], so $R_1$ is the
`defining' representation ($n$) and $R_{n-1}$ is its complex conjugate
$(\bar n)$.}
Quite recently these S-matrices have 
been subjected to a highly stringent
test involving the Thermodynamic Bethe Ansatz
(TBA)---see [\Ref{BNNW}] for the SU($n$) case, and [\Ref{TH3}] for the
other classical groups---using a technique which was first applied to the
O($N$) sigma model [\Ref{HN},\Ref{HMN}] (see also
[\Ref{FNW},\Ref{CGN},\Ref{EH},\Ref{EH1}]) based on ideas developed in
[\Ref{PW},\Ref{W}]. This work showed that the proposed S-matrices for
the PCMs are correct as they stand, without the 
addition of CDD factors, a potential ambiguity which was never resolved in the
original papers. The calculations which were 
performed to test the S-matrices also led, as a by-product, to an exact
formula for the mass gaps of these models, and the results have since
been confirmed by lattice simulations in various cases [\Ref{HAS},\Ref{HH}].

Our aim in this paper is to extend the rather satisfying picture of 
the bosonic principal chiral models sketched above to include their 
supersymmetric versions, at least when the group is $G$ = SU($n$). 
When fermions are coupled to bosonic sigma-models in special 
ways---either minimally or supersymmetrically---integrability can be
maintained at the quantum level, or even re-instated if it was
originally broken through quantum effects in the bosonic theory
[\Ref{AFL}].\note{An example of the latter
phenomenon occurs for the sigma models on ${\Bbb C}P^{n-1}$, or
SU($n$)/SU$(n-1)\times$U(1). }
This is one major motivation for studying supersymmetric integrable
theories. Another point, of more direct relevance to the work
we shall describe here, is that even if the original bosonic theory
{\it is} quantum integrable, the addition of fermions can produce 
dramatic changes, including a radical alteration
of the spectrum. Thus, one finds that typically 
a supersymmetric model involves something `genuinely new', beyond 
the mere addition of superpartners 
for each of the original bosonic states. The original bosonic
states may even disappear completely in some cases.
We shall see below that it is just this kind of radical alteration in
the spectrum which is needed in order to understand 
the supersymmetric SU($n$) PCM.

It is not difficult to construct a supersymmetric counterpart of the 
bosonic lagrangian \BLAG\ for a general group $G$, although there are a
number of slightly different ways to write the end result.
We choose to supplement the original 
bosonic, $G$-valued field $U$ with a 
Majorana fermion $\psi$ taking values in the Lie
algebra of $G$ and we take as the lagrangian\note{This can be derived
in a number of ways; perhaps the most
straightforward is to consider the obvious modification of \BLAG\ in
superspace and then to integrate out the auxiliary fields to arrive at the
expression above.}
$$
{\cal L}={1\over g}{\rm Tr}\left( \, \partial_\mu U^{-1}
\partial^\mu U
- i \bar \psi \gamma^\mu ( \, \del_\mu \psi  + {\textstyle {1\over 2}}
[ U^{-1} \del_\mu U ,
\psi ]
\, ) 
- {\textstyle{1 \over 16}} \{ \bar \psi , \gamma_5 \psi \} 
\{\bar \psi , \gamma_5 
\psi \} \, \right) 
\nfr{SLAG}
(Our conventions for spinors in two dimensions are those of
[\Ref{AFM}].)
Note that the anti-commutators appearing in the fermion interaction
terms are elements of the Lie algebra because of the Grassmann
nature of the fermion fields. 
This model is supersymmetric, and, like its bosonic counterpart,
it has a global chiral symmetry group
$\glr$. In the way we have chosen to write the lagrangian, a general 
element $(g_{\rm L} , g_{\rm R})$ in $\glr$ acts according to 
$$
U \rightarrow g^{\phantom{1}}_{\rm L} U g_{\rm R}^{-1} \, , \quad 
\psi \rightarrow
g^{\phantom{1}}_{\rm R} \psi g_{\rm R}^{-1}
\nfr{CSYM}
which clearly leaves the theory invariant.
It may seem strange at first that the fermions transform only under
$G_{\rm R}$, but this is just a matter of
convention; there is an equivalent way of writing the theory in which
the fermions transform only under $G_{\rm L}$, these 
formulations being related
by a redefinition of $\psi$ by conjugation with $U$.\note{From the
superspace point of view this ambiguity corresponds to a choice
in how one defines the component fields of a given group-valued
superfield.}

>From now on we shall specialize to the case $G$ = SU($n$). In the 
next section we shall construct an S-matrix which we claim describes
the scattering of particles in this theory, based on a certain
assumption about the spectrum.
We shall afterwards 
subject this S-matrix---and hence also our assumption about the
spectrum---to the same stringent test involving the TBA that has already been
successfully carried out in the bosonic case. In this way we shall
be able to show that in this model too, there are no allowed CDD
ambiguities in the S-matrix, and  
we shall be able to extract an exact expression for the
mass-gap. 
Before embarking on the detailed technical aspects of the construction of the
S-matrix and its
verification, some discussion is in order about
what the spectrum of the theory defined by \SLAG\ is likely to be.

To get some idea of how to proceed it is useful to recall
the relationship of PCMs (whether bosonic or supersymmetric) 
to some other well-known integrable
sigma-models, particularly those with O($N$) symmetry for which the
target spaces are the spheres $S^{N-1}$ with their standard round metrics. 
Notice that the first non-trivial member of this series is the O(3)
model, which has an alternative description as a sigma-model with
target space ${\Bbb C}P^1$, reflecting the relationship between
O(3) and SU(2) (the latter group acting naturally as the symmetry group
of complex projective space). The next member of the sequence
is the O(4) model, which is nothing but a re-writing of the SU(2) PCM,
again reflecting the usual homomorphism from SU(2) $\times$ SU(2) to
SO(4). 

The bosonic O($N$) model has a very simple spectrum consisting of a
single degenerate multiplet of particles transforming in the vector
representation of the group, and 
the S-matrix for these states was found for $N \geq 3$
in the original work of [\Ref{ZZ}]. The super O($N$) model, on the
other hand, was
considered in [\Ref{WIT},\Ref{SW}] and was found to have a much richer
spectrum with additional particles transforming in each of the
anti-symmetric tensor
representations of O($N$).
Witten and Shankar determined the S-matrix for the supermultiplet 
transforming in the vector representation of O($N$) when $N > 4$,
but they found no consistent solution for the cases $N = 3, 4$.
The explanation which they proposed was based on the suggestion 
that the super O($N$) model should actually contain yet more
particles: in fact it is not difficult to see that the theory
contains additional kink-like states, transforming in the spinor
representations of O($N$), which arise just as in the Gross-Neveu
model [\Ref{GNK}] 
(the Gross-Neveu model is, after all, the fermionic sector of the 
super O($N$) theory). 
The particles in the vector or tensor representations can
be regarded as bound states of these kinks, and it is now 
natural to suppose that in the particular cases with symmetry
O(3) and O(4) the bound states simply disappear, leaving only the kinks in the
spectrum 
(just as the `elementary' particles in the sine-Gordon
spectrum disappear for sufficiently large coupling, leaving only the solitons).
The S-matrix for the kinks is related to the S-matrix of the kinks of
the supersymmetric sine-Gordon model [\Ref{BL},\Ref{ABL},\Ref{AHN2}] 
(see also [\Ref{HM}]). 
For $N=3$, this picture is confirmed by the equivalence 
with the ${\Bbb C}P^1$ model,
in which the only states transform as doublets of SU(2), ie.~spinors
of O(3). For $N=4$ we are led to conclude that the only
states of the SU(2) PCM should transform as spinors of O(4), or in
other words as representations (1/2,0) and (0,1/2) of SU(2)$\times$SU(2).

This picture of the super O($N$) model advanced by Witten and Shankar 
suggests a natural conjecture concerning 
the spectrum of the super SU($n$) PCMs, based on the coincidence of
these theories for $N=4$ and $n=2$.
Thus, for general $n$, we may expect to find states in the super
SU($n$) PCM transforming in the representations 
$(R_a,1)$ and $(1,R_a)$ of SU($n$)$\times$SU($n$) (where 1 denotes the
trivial representation) rather than 
in the `diagonal' representations $(R_a , \bar R_a)$ found in the
bosonic theory. 
In the bosonic PCM we can regard as `fundamental' the states 
transforming in a representation $(R_1, \bar R_1)$, since all other S-matrix
elements can be deduced from these via the bootstrap equations.
In the supersymmetric theory, we can regard as
similarly fundamental the 
kink degrees-of-freedom transforming in representations 
$(R_1,1)$ and $(1,R_1)$, from which all other S-matrix elements are
determined.
Despite the different representations of the
global symmetry group which appear in the bosonic and supersymmetric
cases, we shall see that the mass relations
\MASS\ are unchanged.

In addition to suggesting the foregoing conjecture for the
spectrum, the coincidence of the O(4) model and the SU(2) PCM also
gives important insight into the way in which supersymmetry enters the 
construction.
A closer examination of the S-matrix for the spinor particles in the
O(4) model [\Ref{HM}] shows that the supersymmetric part is
identical to the soliton S-matrix in the supersymmetric
sine-Gordon theory [\Ref{AHN2}] at a particular choice of the coupling
where the scattering is reflectionless. This S-matrix 
is well-known to be related to an affine SU(2) quantum
group symmetry with deformation parameter $q$ a root of unity corresponding
to level 2 [\Ref{BL}].  
The supersymmetry acts on the kink-quantum numbers in a rather involved 
way that is intimately related to the quantum group
structure, as discussed in [\Ref{QG}].
For the SU($n$) PCM it is sensible to try to implement
supersymmetry in an analogous way, by association with an affine
SU($n$) quantum group with $q$ a root of unity corresponding to 
level $n$. 
We shall have a number of comments to make below regarding 
various subtleties involved
in this construction. 

\chapter{The exact S-matrix}

In this section we recall some general ideas which are
useful in the construction of exact S-matrices. After setting up a
certain amount of 
technology we will be able to give a succinct statement of 
our conjecture for the S-matrix of the supersymmetric SU($n$) PCM.
The main idea in the approach we shall follow here is the notion of an S-matrix
`block' which is invariant under the action of a quantum group
related to SU($n$).\note{The relation of
S-matrices to quantum groups was first developed for SU(2) in
[\Ref{L1}] and later extended to SU($n$) in [\Ref{TH1}]. The extension
to other algebras and to the RSOS representations appears in
[\Ref{TH2}]. A more unified approach was developed in 
[\Ref{D1}].}
Our S-matrix will eventually be constructed in terms
of such blocks and, along the way, we will explain how the S-matrix for the
bosonic PCM can be understood from the same point-of-view.
There are actually two kinds of blocks, one kind 
associated to the vertex-type realization and another associated to the 
RSOS-type representations of a 
quantum group and we shall need both kinds.\note{This language is 
borrowed from integrable lattice
models where the S-matrix elements play the r\^ole of Boltzmann weights} 
The detailed construction of the blocks has been discussed at
length in the papers cited above and so we shall concentrate here on the
characteristic properties of the blocks, giving
only those explicit expressions 
that we shall need later. 

In the vertex picture, 
particle states transform in the fundamental
representations $R_a$, $a=1,\ldots,n-1$, of SU($n$). 
We denote single-particle states by vectors 
$\bxi^{(a)}(\theta)$ living in the vector space $V_a$ which carries
the representation $R_a$, where $\theta$ is the rapidity.
The corresponding blocks are matrices of the form 
${\widetilde S}_{ab}(\theta)_M^N$ 
with $M = ( \bxi^{(a)} , \bxi^{(b)} )$
and $N = ( \bta^{(b)} , \bta^{(a)} )$
which give the amplitude for scattering between particle states
transforming under the given representations:
$$
{\widetilde S}_{ab}(\theta)_M^N:\ \bxi^{(a)}(\theta_1)+\bxi^{(b)}(\theta_2)
\rightarrow\bta^{(b)}(\theta_2)+\bta^{(a)}(\theta_1),
\efr
with $\theta=\theta_1-\theta_2$. 
The blocks enjoy the following properties.
(i) They satisfy the Yang-Baxter equation. (ii) They obey the 
the completeness or unitarity relation
$$
{\widetilde S}_{ab}(\theta){\widetilde S}_{ba}(-\theta)=I.
\nfr{UNI}
(iii) They are crossing symmetric.
(iv) They satisfy 
the $su(n)$ bootstrap equations: if $V_c\subset V_a\otimes V_b$ (so
$c=a+b$ if $a+b<n$ and $c=a+b-n$ if $a+b>n$) then 
$$
{\widetilde S}_{dc}(\theta)\simeq{\widetilde S}_{da}(\theta-iu_b)
{\widetilde S}_{db}(\theta+iu_a),
\nfr{BOOT}
where $u_a=a\pi/n$ if $a+b<n$ and $\pi-a\pi/n$ if $a+b>n$. The
equality in \BOOT\ holds in the sense that the right-hand-side should be
restricted to the subspace $V_c\subset V_a\otimes V_b$. 
In fact we shall take the blocks to be the {\it minimal\/} expressions with
this
property, and we shall explain more fully below what this term means.
(v) Finally, there is actually a one-parameter family of blocks meeting
all these conditions, and we can distinguish between its members by
labelling them with a certain real number $\lambda$.
It turns out that these S-matrix blocks are invariant under the action
of a quantum group symmetry $U_q (su(n)^{(1)})$ where
$$
q=-\exp -i\pi\lambda
\efr 
is the deformation parameter.

Let us elaborate on the last of these properties first.
The quantum group $U_q (su(n)^{(1)})$ is defined by a set of
generators $H_i$ and $E_i^\pm$ with 
$i=0,\ldots,n{-}1$ which obey the `deformed' commutation relations
$$
[H_i , E^\pm_j] = \pm a_{ij} E^\pm_j, \qquad [E^+_i , E^-_j] = 
\delta_{ij} {q^{H_i} - q^{-H_i} \over q - q^{-1}},
\nfr{DCR}
where $a_{ij}$ is the usual Cartan matrix for $su(n)^{(1)}$.
We are interested here in the {\it centerless\/} extension of the
finite-dimensional quantum group $U_q (su(n))$, in other words, 
we will always understand 
$U_q (su (n)^{(1)})$ to mean the quantum loop group. 
The precise way in which this acts as a symmetry of the
S-matrix blocks is rather
subtle, because it incorporates the rapidity operator of the theory in
a non-trivial way. 

To explain how this works, let us begin by introducing the matrices
$\tilde H_i$ and $\tilde E^\pm_i$ $i = 1, \ldots , n{-}1$ which
represent the Cartan generators and simple-root step operators for the
finite-dimensional algebra $U_q (su(n))$ in the representation $R_a$. 
By definition, these act on the
vector space $V_a$. Let us also introduce matrices $\tilde E_0^\pm$
corresponding to the step operators for the (non-simple) lowest root
of $su(n)$, and $\tilde H_0 = - \sum_{i = 1}^{n-1} \tilde H_i$. In
terms of these matrices, the quantum loop-group
generators are realized on single-particle states in $V_a$ with definite
rapidity $\theta$ by 
$$
H_i \rightarrow \tilde H_i \ , \qquad E^\pm_i \rightarrow e^{\pm s_i
\theta} \tilde E^\pm_i
\efr
with $i = 0, \ldots , n{-}1$. Notice that on a fundamental
representation the deformed commutation relations \DCR\ reduce to
those of the Lie algebra because the eigenvalues of the $H_i$ are only $0$
or $\pm1$. The set of real numbers $s_i$
are the Lorentz spins of the symmetry generators, and they can be
viewed as a choice of (non-integral) gradation of the loop algebra.
For the vertex-type S-matrix blocks that we shall write down below, the
appropriate choice for these spins is 
$$
s_i =\lambda , \qquad i = 0 , \ldots , n-1,
\nfr{PGRAD}
so their values are fixed by the
deformation parameter $q$. This can also be expressed by saying that
the loop algebra $U_q(su(n)^{(1)})$ is realized in the principal
gradation, but with the standard loop parameter given by
$$
t = (-q)^{-\theta/i\pi} = e^{\lambda \theta}, 
\efr 
which relates it to the rapidity and to $q$.

Lastly, we should note that to state precisely what is meant by a
symmetry of the S-matrix, we need to extend the action of the
generators introduced above to multi-particle states.
For this we use the coproduct structure in the quantum group.
Thus, the 
generators extend to two-particle states according to 
$$\eqalign{
\Delta(E^\pm_i)& = E_i^\pm\otimes q^{-H_i/2}+q^{H_i/2}\otimes E_i^\pm,\cr
\Delta(H_i)& = H_i\otimes 1+1\otimes H_i.\cr
}
\efr
Invariance under the quantum group means that these generators 
commute with the action of the S-matrix blocks. 

Having explained something of the symmetry properties of the
vertex-type S-matrix
blocks, let us return
to the bootstrap condition \BOOT\ written above.
The blocks are minimal solutions of these crucial equations
in the sense that they have no poles
on the physical strip for values of the parameter in the range
$0<\lambda<1/n$.
This means that although they satisfy the $su(n)$ bootstrap
equations, there are no simple poles to 
signal {\it dynamically\/} that $V_c$ should be a bound-state
of $V_a$ and $V_b$ if $V_c\subset V_a\otimes V_b$. (When we come to
use the blocks to write down physically meaningful S-matrices, the
necessary poles will be incorporated in 
additional scalar factors.)
Nevertheless, the usual 
bootstrap procedure can still be used to build up {\it all\/} the
blocks, for any given representations, out of the elementary
block ${\widetilde S}_{11}(\theta)$ for the scattering of the $n$-dimensional
defining representation $R_1$ with itself. To complete our description
of the blocks for the vertex representations, let us finally give some
explicit formulas.

First of all, it is convenient to label the states by the
weight vectors of the representations. In order to do this we introduce 
the weights ${\bfmath e}_i$, 
$i=1,\ldots,n$ of the representation $R_1$, with inner 
products ${\bfmath e}_i\cdot {\bfmath e}_j=\delta_{ij}-1/n$. Now we can 
write down an expression for the scattering of states
$$
{\widetilde S}_{11}(\theta)_{ij}^{kl}:\ 
{\bfmath e}_i(\theta_1)+
{\bfmath e}_j(\theta_2)\rightarrow{\bfmath e}_k(\theta_2)+{\bfmath
e}_l(\theta_1),
\efr
by specifying the three non-zero components
$$\eqalign{
{\widetilde
S}_{11}(\theta)_{ii}^{ii}&=f(\theta)\sin\left(\pi\lambda-n\lambda\theta/2i
\right)\cr
{\widetilde
S}_{11}(\theta)_{ij}^{ji}&=f(\theta)\sin\left(n\lambda\theta/2i\right),
\ i\neq j\cr
{\widetilde
S}_{11}(\theta)_{ij}^{ij}&=f(\theta)e^{(2j-2i\pm n)\lambda\theta/2}
\sin\left(\pi\lambda\right),\ i>j\ {\rm or}\ i<j.\cr}
\nfr{FB}
The function $f(\theta)$ is again {\it minimal\/} in the sense of 
having the least number of poles and zeros on the physical 
strip---in this case 
none---necessary in order to ensure unitarity and crossing symmetry. The
explicit form of the function is [\Ref{TH1}]
$$\eqalign{
f(\theta)&={1\over\sin(\pi\lambda-n\lambda\theta/2i)}
\prod_{j=1}^\infty
{\Gamma\left(1+{in\lambda\theta\over2\pi}+(j-1)n\lambda\right)\over
\Gamma\left(1-{in\lambda\theta\over2\pi}+(j-1)n\lambda\right)}\cr
&\times{\Gamma\left({in\lambda\theta\over2\pi}+jn\lambda\right)
\Gamma\left(-{in\lambda\theta\over2\pi}+[(j-1)n+1]\lambda\right)
\Gamma\left(1-{in\lambda\theta\over2\pi}+(jn-1)\lambda\right)\over
\Gamma\left(-{in\lambda\theta\over2\pi}+jn\lambda\right)
\Gamma\left({in\lambda\theta\over2\pi}+[(j-1)n+1]\lambda\right)
\Gamma\left(1+{in\lambda\theta\over2\pi}+(jn-1)\lambda\right)}.\cr}
\efr
The information we have just given now determines all the blocks,
because of the bootstrap property.
Of course the scattering matrix of the
charge conjugate states, transforming as $R_{n-1}$, can also be found by
using crossing symmetry. If we introduce a basis for these
states denoted ${\bfmath e}_{\bar i}(\theta)$ then 
we must have
$$
{\widetilde S}_{1,n-1}(\theta)_{i\bar j}^{\bar kl}=
{\widetilde S}_{11}(i\pi-\theta)_{ki}^{lj}.
\nfr{CS}
It is important that the relation \CS\ is consistent with the
bootstrap equations \BOOT\ and also with unitarity \UNI. 

We must now draw attention to some special behaviour of the quantum
group symmetry for various values of $q$.
For generic values, the only part of the conventional SU($n$) symmetry which
survives unscathed in the quantum group is the Cartan
subalgebra. 
But if $q^2=1$ then the quantum group invariance becomes a conventional Lie
algebra invariance, and then the blocks that we have introduced above are
simply SU($n$)-invariant.
Note in particular that the 
S-matrix blocks have a well-defined limit 
as $\lambda\rightarrow0 $, which means
$q\rightarrow-1$; this is known as the `rational' limit. 
The Lorentz spins $s_i$ of the generators become zero in this
limit,
as one would expect. 
The resulting SU($n$)-invariant 
blocks have a special role to play, and 
we shall denote them by 
${\widetilde S}_{ab}^{{\rm SU}(n)}(\theta)$.

The other significant special values 
for $q$ occur when $\lambda=1/(n+k)$ where $k$ is a
positive integer. In these cases it is possible to `restrict' the
vertex-type representations that we have considered up till now and to
pass to the so-called RSOS version of 
the S-matrix, which is realized in terms of kink-like states.
We now need to describe this in some detail.

We must first specify the set of `kink' states on which the RSOS S-matrices
act, and this depends on the choice of the integer $k$.
By definition, the `kinks' interpolate between `vacua' which are
labelled by the set of
integrable highest weights of the affine algebra $su(n)^{(1)}$ at
level $k$. This set of highest weights can be written explicitly as 
$$
\Lambda=
\left\{\sum_{i=1}^{n-1}m_i{\bfmath e}_i,\ k\geq m_1\geq m_2\geq\cdots
\geq m_{n-1}\geq0\right\}.
\nfr{VAC}
where ${\bfmath e}_i$ are the weight vectors introduced earlier.
Now the kinks also come in a number of different `species' which are
labelled by the fundamental representations $R_a$, so we denote them 
$K^{(a)}_{\bmu\bnu}(\theta)$. Here $\bmu,\bnu\in\Lambda$ and the kink can
be thought of as carrying a topological charge $\bnu-\bmu$ which is 
a vector of eigenvalues of the Cartan generators. The
final restriction on the allowed single-kink states is that this
charge must be a weight of the representation $R_a$; namely 
$\bnu-\bmu=\sum_{p=1}^a{\bfmath e}_{i_p}$ with all the $i_p$'s distinct.
This restriction can be
expressed in terms of the representation theory of the 
affine algebra $su(n)^{(1)}$ at level $k$ by saying that there is an
allowed kink state $K^{(a)}_{\bmu\bnu}(\theta)$ whenever the irreducible
module with highest weight $\bnu$ occurs in the decomposition of the
tensor product between the module with highest weight $\bmu$ and the
module associated to $R_a$.
What we have said fixes the allowed one-kink states completely. Multi-kink
states are now constructed as usual in quantum mechanics, except that
there is an additional `adjacency' condition which requires 
that successive vacua
must coincide. Thus, the allowed two-kink states, for example, are of
the form $ | K_{\bmu_1\bmu_2}^{(a_1)}(\theta_1)
K_{\bmu_2\bmu_3}^{(a_2)}(\theta_2) \rangle $ and similarly for any
number of kinks. 

We can now introduce the idea of an S-matrix block associated to the
scattering of these RSOS kinks, just as we did previously for the
vertex representations.
These RSOS blocks share many of the nice properties of the vertex
blocks that we discussed
earlier: (i) They satisfy the Yang-Baxter equation; (ii) they obey the
unitarity condition; (iii) they are crossing symmetric; 
(iv) they obey the $su(n)$ bootstrap equations.
The bootstrap property means that we can, if we so wish, 
consider only the S-matrix 
elements for the kinks associated to the representation $R_1$, since
all the others may be deduced from these.
The S-matrix elements 
$$\eqalign{
{\widetilde S}_{11}^{({\rm RSOS},k)}(\theta;\bmu)_{ij}^{kl}:\ &
K_{\bmu,\bmu+{\bfmath e}_i}^{(1)}(\theta_1)+
K_{\bmu+{\bfmath e}_i,\bmu+{\bfmath e}_i+{\bfmath e}_j}^{(1)}(\theta_2)\cr
&\rightarrow
K_{\bmu,\bmu+{\bfmath e}_k}^{(1)}(\theta_2)+
K_{\bmu+{\bfmath e}_k,\bmu+{\bfmath e}_k+{\bfmath e}_l}^{(1)}(\theta_1),\cr}
\efr
are, like those in \FB , non-zero only if charge is conserved,
i.e. ${\bfmath e}_i+{\bfmath e}_j={\bfmath e}_k+{\bfmath e}_l$. 
The explicit expressions for the non-zero
elements are
$$\eqalign{
{\widetilde
S}_{11}^{({\rm RSOS},k)}(\theta;\bmu)_{ii}^{ii}&=f(\theta)\sin\left(\pi\lambda-
n\lambda\theta/2i\right)\cr 
{\widetilde
S}_{11}^{({\rm RSOS},k)}(\theta;\bmu)_{ij}^{ji}&=f(\theta)
\sin\left(n\lambda\theta/2i\right){\sqrt{\left(s_{ij}(\bmu+{\bfmath e}_i)s_{ij}
(\bmu+{\bfmath e}_j)\right)}\over
s_{ij}(\bmu)},\ i\neq j\cr
{\widetilde
S}_{11}^{({\rm RSOS},k)}(\theta;\bmu)_{ij}^{ij}&=f(\theta)\left[
\sin\left(\pi\lambda-n\lambda\theta/2i\right)
+\sin\left(n\lambda\theta/2i\right){s_{ij}(\bmu+{\bfmath e}_i)
\over s_{ij}(\bmu)}\right],\ i\neq j\cr}
\nfr{FR}
(compare with \FB ) where
$$
s_{ij}(\bmu)=\sin\left(\pi\lambda({\bfmath e}_i-{\bfmath 
e}_j)\cdot(\bmu+\brho)\right),
\efr
and $\brho=\sum_{j=1}^{n-1}(n-j){\bfmath e}_j$ 
is the sum of the fundamental weights of $su(n)$.

The last point to discuss, which is central to our construction, is
the issue of the symmetry properties of these RSOS blocks. 
We explained above that the vertex blocks are invariant under a 
quantum loop group $U_q (su(n)^{(1)})$ where the generators have
non-trivial Lorentz spins given by \PGRAD . 
The first step in the construction of the RSOS blocks involves starting 
with the vertex S-matrix and carrying out a conjugation operation 
which has the
effect of changing the action of the $U_q(su(n)^{(1)})$ symmetry
generators so that they have modified spins, with \PGRAD\ replaced by
$$
s_0 = n \lambda , \qquad s_i = 0, \ i = 1 , \ldots, n{-}1.
\nfr{HGRAD}
This is effectively a change to the homogeneous gradation of the loop
algebra. 
In a loose sense, the RSOS reduction is now achieved by 
`modding-out' with respect to the generators which have just been rendered
spin-less by this change in gradation 
(more properly the RSOS blocks appear as intertwiners of irreducible
representations of the finite-dimensional $U_q(su(n))$) 
so there is now just one surviving conjugate pair of symmetry charges
with non-zero spin. 
The last step, which requires $\lambda = 1/(n + k)$, is to further
restrict the allowed space of states (passing from SOS to RSOS
in statistical mechanical terminology) according to the rules for
allowed kinks which we expressed above in terms of the allowed highest
weights of $su(n)$ at level $k$. 
So we arrive finally at a set of S-matrix blocks for which most
of the original quantum group symmetry has been lost, but for which there
remains a pair of conserved charges with Lorentz spins
$$
\pm s_0 = \pm n \lambda = \pm {n \over n + k}.
\efr

At this stage the possible connection with supersymmetry becomes
apparent: on choosing $k = n$ the formula above tells us that the surviving
charges have Lorentz spin $\pm 1/2$, and it is natural to suspect that
they are supersymmetry generators.
Unfortunately, it is apparently not known at present how to  
show directly that these conserved quantities really obey the correct
supersymmetry algebra for all values of $n$.
The question of how the original quantum group relations descend to
the RSOS picture seems to be rather poorly understood, and there is even
some considerable freedom in how these conserved
charges are defined to act in the restricted theory. (See the discussions in
[\Ref{BL},\Ref{ABL},\Ref{AHN2},\Ref{QG},\Ref{LNW}].)
What has been known for certain for some time 
is that in the simplest case, $n = 2$, it is possible to show
explicitly that 
spin-1/2 conserved quantities can be defined so as to obey the supersymmetry
algebra [\Ref{BL}]. 
We will make the reasonable assumption that this can also be
done for general values of $n$. We have carried out a partial check of
the next simplest case based on SU(3) using brute force methods, and
it appears that in
this case too the residual spin half charges can be defined so as to
obey the desired algebra. In the absence of a general
construction for SU($n$), however, we defer a more detailed discussion of this
point to another occasion. 

To summarize: we will proceed on the assumption that the RSOS
blocks with $k = n$ carry a good representation of supersymmetry.
To add some additional reassurance on this point, we should also 
emphasize that the S-matrix which we write down
based on this assumption will 
ultimately be subjected to a very stringent test.
The conclusion will be that it does indeed correspond to the original
supersymmetric lagrangian \SLAG .

We have now set up all the S-matrix technology that we need, but 
before we consider the supersymmetric model it may be helpful to  
explain how the
S-matrix of the bosonic PCM is constructed from the point-of-view we
have followed here. For the bosonic SU($n$) model, the 
particles transform
in multiplets $(R_a , \bar R_a) = (R_a,R_{n-a})$ of the global 
${\rm SU}(n)_{\rm L}\times{\rm SU}(n)_{\rm R}$ symmetry. 
The states in the theory are consequently of the form  
$$ \vert \bxi^{(a)} , \bta^{(n-a)} ; \theta \rangle
\efr
The S-matrix acting on these states is constructed out
of two SU($n$)-invariant vertex blocks:
$$
S_{ab}(\theta)=X_{ab}(\theta)\, 
{\widetilde S}_{ab}^{{\rm SU}(n)}(\theta)\otimes
{\widetilde S}_{n-a,n-b}^{{\rm SU}(n)}(\theta),
\efr
where the tensor product structure corresponds to the product
structure of the states in the obvious way. 
Notice that there are no reflection processes between the degenerate
multiplets $(R_a,R_{n-a})$ and
$(R_{n-a},R_{a})$. The pre-factors
$X_{ab}(\theta)=X_{n-a,n-b}(\theta)$ are a set of scalar functions 
which obey the SU($n$) bootstrap equations,
the unitarity condition and crossing symmetry independently, but which
introduce simple poles on the
physical strip at just the right positions in order that the particle
$c$ appears as a bound-state of
$a$ and $b$ if $c=a+b$ or $c=a+b-n$. These factors make the bootstrap
dynamical. 
In fact $X_{ab}(\theta)$ is simply the minimal purely elastic S-matrix
associated to $su(n)$:
$$
X_{ab}(\theta)=\prod_{j=\vert a-b\vert+1\atop{\rm step}\ 2}^{a+b-1}
{\sin(\theta/2i+\pi(j-1)/2n)\sin(\theta/2i+\pi(j+1)/2n)\over
\sin(\theta/2i-\pi(j-1)/2n)\sin(\theta/2i-\pi(j+1)/2n)}.
\nfr{MIN}
(In comparing this to the result of [\Ref{ORW}] one should note that 
our definitions of the group SU$(n)_{\rm R}$ differ by complex
conjugation.) 

We are now ready to write down our conjecture for the states and 
scattering amplitudes in the  
supersymmetric SU($n$) PCM. The two essential global invariances which we need
to incorporate are ${\rm SU}(n)_{\rm L}\times{\rm SU}(n)_{\rm R}$ and
supersymmetry. The first of these can be built in using the
vertex-type blocks introduced above, once we have decided what the
allowed representations should be. Following our discussion in the
first section which compared PCMs to the family of O($N$) sigma models,
we conjecture that the super PCM has a spectrum of particles  
with masses $m_a$, given by \MASS , which transform in reducible
multiplets $(R_a,1)\oplus(1,R_a)$ for $a = 1 , \ldots , n{-}1$. 
(Of course, such states are always 
degenerate with their conjugates which transform as 
$(R_{n-a},1)\oplus(1,R_{n-a})$.) 		
To incorporate supersymmetry, these states must also carry additional
quantum numbers on which the super-charges act. Following our
discussion above of RSOS-type blocks and their symmetry properties, we
shall take these to be RSOS kinks of type $a$ with $k=n$.
In short then, the states in the model are of two general types 
$$
{\rm L} \, : \,  \vert\bxi^{(a)},0,K^{(a)}_{\bmu\bnu};\theta\rangle
\qquad 
{\rm R} \, : \,  \vert0,\bxi^{(a)},K^{(a)}_{\bmu\bnu};\theta\rangle
\nfr{SPEC} 
where SU$(n)_{\rm
L}$ acts on the first quantum number, ${\rm SU}(n)_{\rm R}$ acts on
the second quantum number, and supersymmetry acts on the kink degrees 
of freedom. 
We emphasize that the allowed kink states interpolate between the 
integrable weights of $su(n)$ at level $n$ according to the rules
which we summarized earlier. 

Having specified the detailed struture of the states, we define the 
S-matrix as follows. 
The scattering between the L multiplets is 
given by
$$
S_{ab}^{\rm LL}(\theta)=X_{ab}(\theta) \, {\widetilde S}_{ab}^{{\rm
SU}(n)}(\theta)\otimes I \otimes{\widetilde S}_{ab}^{({\rm RSOS},n)}(\theta),
\nfr{LLS}
and similarly between the R multiplets
$$
S_{ab}^{\rm RR}(\theta)=X_{ab}(\theta) \, I \otimes {\widetilde S}_{ab}^{{\rm
SU}(n)}(\theta)\otimes{\widetilde S}_{ab}^{({\rm RSOS},n)}(\theta)
\nfr{RRS}
where $X_{ab} (\theta)$ is the minimal elastic factor defined by \MIN .
The scattering between the L and R multiplets is defined to be 
$$
S_{ab}^{\rm LR}(\theta)= I \otimes I \otimes 
{\widetilde S}_{ab}^{({\rm RSOS},n)}(\theta).
\efr
As usual, the tensor products are to be understood with respect to the
product structure of the states exhibited in \SPEC .
We emphasize that this is a conjecture for the {\it complete\/}
S-matrix, describing scattering amongst {\it all\/} the states in the
theory.

The following points are worthy of note. (i) The scattering between
the L and
R multiplets is completely diagonal in the space of global quantum
numbers, as required by the form of the global symmetry. (ii) It is
only the LL or RR scattering which
lead to bound-states, because only these elements have poles on the
physical strip provided by the scalar factors $X_{ab}(\theta)$. The
LR scattering elements have no poles on the physical strip so
no bound-states which transform non-trivially under both SU$(n)_{\rm L}$ and
SU$(n)_{\rm R}$ are formed. 
(iii) Since the S-matrix elements are built out of the
blocks, we can be assured that all the S-matrix axioms are
satisfied. Furthermore, due to the existence of simple poles on the
physical strip in just the right positions, provided by the factors
$X_{ab}(\theta)$, the multiplets can all be considered as
bound-states of the multiplets with $a=1$. (v) There are no reflection
amplitudes between any two of the degenerate multiplets $(R_a,1)$,
$(R_{n-a},1)$, $(1,R_a)$ and $(1,R_{n-a})$. 

Thus far, we have offered very little evidence to support our claim that
this is the correct S-matrix for the super SU($n$) PCM.
In the following sections, however, we will carry out a very
substantial test of this proposal. We shall find that our conjecture 
is completely consistent with the lagrangian \SLAG , and in particular that
it correctly reproduces the universal part of the beta-function.
The nature of this test is explained more fully below. At this stage,
however, it may also be helpful to draw attention to one {\it
specific\/} kind of possible ambiguity in the S-matrix
which our test will ultimately resolve.

As it stands, the S-matrix we have written down is the 
minimal\note{`Minimality' in this context means the expression
with the smallest number of poles and zeros on the physical strip\/.}
expression which satisfies all the axioms of S-matrix theory, along with the
requirement that the states can all be formed as bound-states on the
elementary multiplet $(R_1,1)\oplus(1,R_1)$. The latter requirement was
responsible for introducing the factors of $X_{ab}(\theta)$ in 
\LLS\ and \RRS. However, it is always possible to multiply these
expressions by CDD factors which satisfy all the axioms 
independently, which passively respect the bootstrap
equations, and, moreover, introduce no additional poles onto the
physical strip (although they introduce additional zeros). These
factors are of the form
$$
Y_{ab}(\alpha;\theta)=
\prod_{j=\vert a-b\vert+1\atop{\rm step}\ 2}^{a+b-1}
{\sin(\theta/2i-\pi(j-1+\alpha)/2n)\sin(\theta/2i-\pi(j+1-\alpha)/2n)\over
\sin(\theta/2i+\pi(j-1+\alpha)/2n)\sin(\theta/2i+\pi(j+1-\alpha)/2n)},
\nfr{CDD}
where $\alpha$ is a parameter $0<\alpha<2$. In principle, an
arbitrary number of these factors with different $\alpha$'s
could be introduced in the S-matrix elements LL and  RR, and a
different set in LR scattering. 
One of the conclusions we shall reach
in the following sections 
is that there are no CDD factors of this type allowed in the 
S-matrix for the models we are considering.

\chapter{S-matrix versus lagrangian: perturbative calculation}

Having arrived at a candidate S-matrix for the supersymmetric SU($n$)
PCM, our aim is now to test this proposal by comparing it 
with the original lagrangian \SLAG .
There is a technique for doing this which is by now well-established
and which has been applied to a number of
different models, following the original pioneering work
of [\Ref{HN},\Ref{HMN}] for the O($N$) sigma-model, so our 
explanation of the general method will be rather brief.\note{A 
recent summary which attempts to collect together the known results was
given in [\Ref{EH4}]; a more complete account is planned for the
near future [\Ref{EH3}]. }
The technique involves modifying the Hamiltonian of the theory 
${\cal H}\rightarrow {\cal H}-hQ$ where $Q$ is
a conserved charge which generates a global symmetry of the model---in
our case a generator of $\sunlr$---and $h$ is a 
parameter with the dimensions of mass which we put in by hand. 
The idea is to calculate the ground-state energy of this new
Hamiltonian, both in perturbation theory and from the S-matrix using
the TBA, and then to compare the results to test 
the lagrangian/S-matrix equivalence.
In fact neither of these calculations can be performed exactly, but 
we can develop asymptotic expansions for the results which 
are assumed to be valid when $h$ is very large and which are 
sufficient to give a highly non-trivial consistency check. 

A standard perturbative analysis of the model \SLAG\ shows that it is
asymptotically free, with the coupling constant behaving as
$$
{1 \over g(\mu )} = 
\beta_1 \ln {\mu \over \Lambda} + {\beta_2 \over \beta_1} \ln
\ln {\mu \over \Lambda} + {\cal O} \left(\ln \ln {\mu \over \Lambda}
{\Big /}
\ln
{\mu \over \Lambda} \right)
\efr
where the universal part of the beta function is given by the coefficients
$$
\beta_1 = n / 8\pi , \quad \beta_2 = 0 .
\efr 
The scale $\Lambda$ in these equations is defined by the requirement
that there is no constant term in the expansion for the running
coupling written above; this depends on 
the renormalization prescription being used and we shall eventually
specialize to the MS-bar scheme.
For a given choice of the charge $Q$ introduced above, it is not
difficult to calculate the new ground-state energy to one-loop,
or in principle to some higher order in the loop expansion,
and when $h\gg\Lambda$ we can 
obtain an expression for the change in the ground-state energy density:
$$\eqalign{ \delta {\cal E}(h)& = {\cal E}(h)-{\cal E}(0)
=h^2 f_1 (h /\Lambda), \cr
f_1 (h  / \Lambda) & = 
{a_0 \over g(h) } + a_1 + {\cal O}(g(h )) \cr
}
\efr
where the numbers $a_j$ are dimensionless quantities resulting from a 
calculation at a certain number of loops indicated by their
subscripts.
Notice that the argument of the coupling constant in this expression is
$h/\Lambda$, which is a consequence of the fact that the final result
must be a renormalization-group invariant quantity and therefore
independent of $\mu$, the subtraction scale.

In the next section we shall see how a similar expression can be found
from the S-matrix via the TBA, but rather than being a function of 
$h/\Lambda$ the result is then of the form
$$\delta {\cal E}(h) = {\cal E}(h)-{\cal E}(0)
=h^2 f_2 (h / m ) 
\efr
where $m$ is the physical mass of the particle states.
Equating this to the previous expression we see that 
$f_1 (h / \Lambda) = f_2 (h
/ m)$ which gives a stringent test of the S-matrix and, if we know 
each expression to sufficient accuracy, this equality 
will allow us to extract the mass-gap $m/\Lambda$.
A simple but very important observation is that the nature of the
functions 
$f_1$ and $f_2$, and hence the values of the constants
$a_j$, is dependent on the choice of the charge $Q$.
It turns out that if we choose $Q$ so that $a_0 \neq 0$, in other words
so that there is a tree-level contribution to the ground-state energy,
then it is sufficient to carry out a one-loop perturbative calculation
in order to get a convincing test of the S-matrix and to extract the
mass-gap. If $a_0 =0$ on the other hand, then one must work harder 
to obtain a non-trivial check, and the mass-gap can only be obtained
by calculating to three-loops or beyond.

It is clearly advantageous to choose a charge $Q$
which leads to one-loop computation and there is actually a general 
strategy by which this can be accomplished for 
a wide 
class of (super) sigma-models [\Ref{EH3}].
In this paper, however, we shall simply mimic the choice of charges 
already considered in the bosonic case [\Ref{BNNW}]
and show that these lead to the desired classical term $a_0 \neq
0$, so that a one-loop calculation will suffice for a test of the
supersymmetric models too.
To be specific, we shall consider here a modification of the
Hamiltonian 
$$ {\cal H} \rightarrow {\cal H} - h Q \qquad {\hbox{\rm where}} \qquad
Q = (\bq\cdot\bH, - \bq\cdot\bH)
\nfr{CHARGE}
is a hermitian generator in the Lie algebra
of $\sunlr$, with $\bH$ denoting the vector of
generators in the Cartan subalgebra of $su(n)$ and $\bq$ some vector
whose components are arbitrary numbers at this stage. From 
the action of the $\sunlr$ symmetry on the fields given in \CSYM\ 
we can deduce that the linearized action generated by this choice
of $Q$ is 
$$\eqalign{
\delta U & = i (\bq \cdot \bH \, U + U \bq \cdot \bH ) \cr 
\delta \psi & = - i (\bq \cdot \bH \, \psi - \psi \, \bq \cdot \bH). \cr 
}\efr
The desired change in the Hamiltonian written in \CHARGE\ 
can now be effected by making the 
substitution in the lagrangian \SLAG :
$$\eqalign{
\del_0 U & \rightarrow \del_0 U + i h (\bq \cdot \bH \, U + U \bq
\cdot \bH ), \cr
\del_0 \psi & \rightarrow \del_0 \psi 
- i h (\bq \cdot \bH \, \psi - \psi \, \bq \cdot \bH) \cr 
}
\efr
and we must expand the resulting expression in powers of $h$ and in powers of 
unconstrained fields so that we can identify the contributions which enter to
various numbers of loops.

To carry out such an expansion, we first write the bosonic field in the form 
$$
U = \exp i \left ( \,{\bfmath n} \cdot \bH + \sum_{\balpha > 0} 
( n_{\balpha} E_{\balpha} + n^*_\balpha E^{\phantom{*}}_{-\balpha} ) \right )
\efr 
where $\bfmath n$ (with no subscript) stands for a vector of real fields
associated to our chosen Cartan subalgebra, $n_{\balpha}$ are complex
fields corresponding to the {\it positive\/} roots, and the sum 
extends over just these positive
roots of the finite-dimensional Lie algebra $su(n)$.
It is now a simple exercise to show that there is indeed a tree-level
contribution for any non-zero
choice of the vector $\bq$ and so a one-loop calculation will suffice
for the purposes of making a non-trivial comparison with the S-matrix.
Because of this, we need only
keep terms quadratic in the fields, and it is not difficult to show
that the fermions decouple completely to this order.
We may further simplify the result by discarding all terms independent
of $h$, since we are interested only in how the result changes as a
function of this background parameter.
This implies that the contribution to the change in the ground-state
energy at one-loop is given by a lagrangian 
$$
{\cal L}_{ \rm 1-loop} = { 4h^2\bq^2 \over g} + {1 \over g}
\sum_{\balpha >0}
( \, \del n^{\phantom{*}}_\balpha \del n_{\balpha}^* -  h^2 (\bq 
\cdot \balpha )^2 n^{\phantom{*}}_{\balpha}n^*_\balpha \, )
\efr
which corresponds to a tree-level term, plus a number of free, massive
bosons (whose fields have been re-scaled to give canonical 
normalizations). 
Using standard dimensional regularization with the $\overline{\rm MS}$
subtraction scheme, the result for the change in the energy density as a
function of the running coupling is now found to be
$$\eqalign{
\delta
{\cal E}(h) & =-{4h^2\bq^2\over g(h)}-{h^2\over4\pi}\sum_{\balpha>0}
(\bq\cdot\balpha)^2\left[\ln(\bq\cdot\balpha)^2-1\right]+{\cal
O}( g(h)) \cr
& =-{h^2 \bq^2 n\over 2 \pi } \ln {h \over \Lambda_{\overline{\rm MS}}} 
-{h^2\over4\pi}\sum_{\balpha>0}
(\bq\cdot\balpha)^2\left[\ln(\bq\cdot\balpha)^2-1\right]+{\cal
O}\left( \ln \ln {h \over \Lambda_{\overline{\rm MS}} } {\Big /} 
\ln {h \over
\Lambda_{\overline {\rm MS}}} \right)
}
\efr
where we have simply substituted the 
two-loop expression for the running coupling given earlier in order to
write the answer explicitly as a function of $h$. 

To complete our perturbative calculation, we wish to make
some convenient specific choices for the vector $\bq$ which defines $Q$
in \CHARGE\ and which has so far been left arbitrary.
We recall that the general method we are seeking to apply 
entails a consideration of the states with
the largest charge/mass ratio, since these will make a dominant
contribution to the new ground-state when $h$ becomes sufficiently
large (we will
discuss this in a little more detail in the next section). We will
follow standard convention and agree to normalize $Q$ so that it has eigenvalue
+1 on this preferred set of states. We would like to choose $\bq$ so
that there are as few of these preferred states as possible, since any
reduction in the number leads to a significant simplification in the
TBA calculation.
  
All states in our theory belong to either left-handed or right-handed 
representations $R_a$ of SU$(n)$, and for each of these representations we are
free to choose a basis of
simultaneous eigenstates of $\bfmath H$, the eigenvalues being just the
weights, of course. The action of $Q$ on
these states is then 
$$\eqalign{
Q\vert\bxi^{(b)},0,K^{(b)}_{\bmu\bnu};\theta\rangle
&= \phantom{-} (\bq\cdot\bxi^{(b)})\vert\bxi^{(b)},0,K^{(b)}_{\bmu\bnu};
\theta\rangle , \cr
Q\vert0,\bxi^{(b)},K^{(b)}_{\bmu\bnu};\theta\rangle
&= - (\bq\cdot\bxi^{(b)})\vert0,
\bxi^{(b)},K^{(b)}_{\bmu\bnu};\theta\rangle.\cr}
\efr
The idea now is that the best we can do to select a small number of 
preffered states is to pick out those
corresponding to the highest weight $\bomega_a$ of the representation
$R_a$ of SU$(n)_{\rm L}$, or the lowest weight $-\bomega_a$ of the
representation $R_{n-a}$ of SU$(n)_{\rm R}$ by choosing 
$\bq$ proportional to $\bomega_a$ for some fixed $a$.
More precisely, it is clear that the correct normalization of the
charge $Q$ is achieved
by taking 
$$
{\bfmath q}=\bomega_a/(\bomega_a^2)
\nfr{WEIGHTS} 
and that then the kink-multiplets
$$
\vert\bomega_a,0,K^{(a)}_{\bmu\bnu};\theta\rangle\ \ {\rm and}\ \ 
\vert0,-\bomega_a,K^{(n-a)}_{\bmu\bnu};\theta\rangle
\nfr{STA}
are indeed the states in the spectrum with the largest charge/mass
ratio having $Q$ eigenvalue +1. 
If we substitute the choice of $\bq$ given by \WEIGHTS\ into our general
formula for the ground-state energy given above we find 
$$
\delta {\cal E}(h)=-{h^2n^2\over2\pi
a(n-a)}\left[\ln{h\over\Lambda_{\overline{\rm MS}}}+\ln\left({n\over 
a(n-a)}\right)-{1\over2}+{\cal
O}\left({\ln\ln(h/\Lambda_{\overline{\rm MS}})
\over\ln(h/\Lambda_{\overline{\rm MS}})}\right)\right]
\nfr{FEPT}
which we shall be able to compare with the result of the S-matrix
calculation carried out in the next section.

Before concluding this section it may be useful to make some
comparison with the analogous calculation for the bosonic PCM as analyzed
in [\Ref{TH3},\Ref{BNNW}].
The one-loop lagrangian derived above is exactly the same, 
{\it as a function of\/} $\bq$, as that found in the bosonic case, and 
this means that the one-loop expression for the ground-state energy is 
unchanged, {\it as a function of the running coupling\/} $g$ {\it and
of\/} $\bq$.
Superficially, then, it may seem that our perturbative calculation is 
ignorant of the presence of fermions in the theory, since they 
have decoupled to this order. In fact the fermions do, nevertheless,
play a r\^ole, and the final result for the ground-state energy 
{\it as a function of\/} $h$, is 
different from the bosonic case. This happens for two reasons. 
First, the running coupling behaves quite differently in the theory
with fermions and, specifically, the vanishing of the second $\beta$-function
coefficient is characteristic of a sigma-model with supersymmetry [\Ref{AFM}].
This means that we get a different result for the ground-state energy
{\it as a function of\/} $h$ {\it and\/} $\bq$.
Secondly, there is a further modification because, compared to the
bosonic case, the choice of $\bq$ 
written in \WEIGHTS\ involves a different
normalization. 
This reflects the fact that in the supersymmetric PCM all states lie in 
either left- or right-handed representations while in the bosonic
theory they live in diagonal representations (requiring an extra
factor of two to ensure $Q$ has maximum eigenvalue +1).

\chapter{S-matrix versus lagrangian: TBA calculation}

We now turn to the calculation of $\delta {\cal E}(h)$ from the
S-matrix using the TBA. 
It is clear that if we make the choice of charge given in \WEIGHTS\
and then increase $h$ 
from zero until it exceeds the threshold value $m_a$, 
it will become energetically
favourable to populate the ground-state with particles in the
multiplets \STA. We shall assume in what follows that the new ground-state
effectively contains
{\it only\/} particles corresponding to the preferred states \STA .
The idea behind this assumption is that the preferred states \STA\ with the
largest charge/mass ratio should repel all other states and thereby
dominate the new ground-state.
Some support for this picture is obtained by considering 
which other states might also 
appear in the new ground-state. For it to be energetically
favourable for a state, labelled by $\bxi^{(b)}$, to appear in the
ground-state, it must have positive charge,
i.e.~$\bxi^{(b)}\cdot\bomega_a>0$.
But if this condition is satisfied then
$\bxi^{(b)}+\bomega_a$ cannot be a weight of a fundamental
representation (the latter requires 
$\bxi^{(b)}\cdot\bomega_a<0$ as a necessary condition); 
hence the states labelled by $\bxi^{(b)}$ cannot form bound-states
with the states \STA\ from which we deduce that 
the forces between them cannot be
attractive. This kind of argument should make our assumption seem
plausible,
but it does not provide a rigorous proof.
In fact, our hypothesis {\it can\/} actually be proven from an analysis of the
full TBA equations of the theory, but the proof is rather technical 
and so we have chosen not to reproduce it here 
(see [\Ref{EH3}] for details), relying instead on the less rigorous but more
physical foregoing arguments. The reason the hypothesis is so
important is that 
it allows us to deal directly with a much simpler set of TBA equations
because we can immediately restrict to configurations containing 
only the states \STA. Our final result---the
agreement between the perturbative calculation and the TBA calculation---will
also confirm that this working hypothesis is correct.

The idea behind the TBA is to consider the thermodynamics of 
a gas of the particles which
interact via the exact factorizable S-matrix [\Ref{TBA}]. Since the number of
particles is preserved it makes sense to consider single particle
states. One then imposes periodic boundary conditions to get equations
relating the densities-of-states of the various particles. From these
equations conventional thermodynamic arguments lead to an expression
for the free energy at finite temperature and with chemical potentials.
The ground-state energy density in the presence of a coupling to a
charge corresponds to the zero temperature limit with some specific
chemical potential. The difficulty facing us is that, even though we
have restricted to the subspace of states \STA , these degrees of
freedom still do not interact purely elastically as far as their kink
degrees-of-freedom are concerned. In finding the equations for the particle
densities-of-states one has to perform a diagonalization in the kink
subspace. Fortunately, however, the S-matrix elements in this subspace
are proportional to RSOS Boltzmann weights of an integrable lattice
model and the relevant diagonalization has already been
performed\note{See [\Ref{TH4}] for a discussion in the context of
S-matrix theory. In fact not all the relevant eigenvalues have been
obtained, as far as the authors are aware; however, the resulting TBA
equations are identical to those conjectured for SU($n$) on the basis
of the universality of the TBA equations [\Ref{RM}].} and
we shall simply quote the result for the Bethe equations which relate the
densities-of-states. 

We denote the densities of occupied states in rapidity space 
of the two multiplets in \STA\ as $\sigma_{\rm
L}(\theta)$ and $\sigma_{\rm R}(\theta)$. Our normalization is such
that, for example
$\int_{-\infty}^\infty d\theta\sigma_{\rm L}(\theta)$
gives the number of occupied L-states per unit length of real space.
The densities of un-occupied
states, or holes, are denoted $\tilde\sigma_{\rm L}(\theta)$ and 
$\tilde\sigma_{\rm
R}(\theta)$, respectively, so that the total densities-of-states are therefore
$\tilde\sigma_{\rm L}(\theta)+\sigma_{\rm L}(\theta)$ and
$\tilde\sigma_{\rm R}(\theta)+\sigma_{\rm R}(\theta)$, respectively.

The diagonalization in the
kink subspace introduces additional terms which behave as if they come
from particles with zero mass
associated to the simple roots of SU($n$) and carrying a `string-length'.
These fictitious particles are known as
magnons and we denote their densities as $r^a_p(\theta)$ and their
associated hole densities as $\tilde r^a_p(\theta)$, 
where $p$, the string-length
index, and $a$, which labels the simple roots of the algebra $su(n)$, 
both run from $1$ to $n-1$.\note{In general the string length index
runs from 1 to $k-1$, but in this case the level $k=n$} 
The Bethe equations relating these densities are
$$\eqalign{
\tilde\sigma_{\rm L}(\theta)+B^{(a)}*\sigma_{\rm L}(\theta)
+C^{(a)}*\sigma_{\rm R}(\theta)+a_p^{(n)}* r_p^a(\theta)&=
{m_a\over2\pi}\cosh\theta,\cr
\tilde\sigma_{\rm R}(\theta)+B^{(a)}*\sigma_{\rm R}(\theta)
+C^{(a)}*\sigma_{\rm L}(\theta)+a_p^{(n)}* r_p^{n-a}(\theta)&=
{m_a\over2\pi}\cosh\theta,\cr}
\nfr{BEI}
along with the magnon equations 
$$
\tilde r_p^b(\theta)+A_{pq}^{(n)}* K_{bc}^{(n)}
* r_q^c(\theta)=\delta_{ba}a_p^{(n)}*
\sigma_{\rm L}(\theta)+\delta_{b,n-a}a_p^{(n)}*\sigma_{\rm R}(\theta).
\nfr{BEM}
In \BEI\ and \BEM\ we have used the notation $f*
g(\theta)=\int_{-\infty}^\infty
d\theta'f(\theta-\theta')g(\theta')$ and $a,b$ and $p,q$ all run from
$1$ to $n-1$ (repeated indices are summed). 
The kernels appearing in \BEI\ are
$$\eqalign{
B^{(a)}(\theta)&=\left[A_{nn}^{(\infty)}\right]^{-1}* A_{aa}^{(n)}
(\theta),\cr
C^{(a)}(\theta)&=\left[A_{nn}^{(n+1)}\right]^{-1}
*A_{n-a,a}^{(n)}(\theta)-A_{n-a,a}^{(n)}(\theta),\cr
a_p^{(n)}(\theta)&={1\over2\pi}\cdot{\sin(\pi p/n)\over\cosh\theta-
\cos(\pi p/n)}.\cr}
\efr
If we define Fourier transforms as
$$
f(\theta)=\int_0^\infty{dx\over\pi}\cos(\theta x)
\hat f(x),
\efr
then by definition 
$$
[f]^{-1}(\theta)=\int_0^\infty{dx\over\pi}\cos(\theta x)
{1\over\hat f(x)},
\efr
and to complete the definition of the kernels we have
$$
\hat A_{pq}^{(k)}(x)={2\sinh({\rm min}(p,q)\pi x/n)\sinh((k-{\rm max}(
p,q))\pi x/n)\cosh(\pi x/n)\over\sinh(k\pi x/n)\sinh(\pi x/n)},
\efr
where $1\leq p,q\leq k-1$ and $\hat K_{pq}^{(k)}(x)=\left(\hat 
A^{(k)}(x)^{-1}\right)_{pq}$.
The kernels are related to the S-matrix elements of the states \STA,
but unfortunately not in a simple way.\note{The 
reason for this is that the macro-states are actually
full of magnon $n$ strings and so the kernels $B^{(a)}(\theta)$ and
$C^{(a)}(\theta)$ are the derivatives of the phases shifts of the
scattering of the particles \STA, but in the background of an $n$
string magnon `sea' [\Ref{TH4}].}

To find the TBA equations (in our case at zero temperature), 
in the
presence of the coupling to the charge, one minimizes the value of the 
new Hamiltonian $H-hQ$, which for a macroscopic configuration is
$$
\int_{-\infty}^\infty{d\theta\over2\pi}\left(m_a\cosh\theta-h\right)
\left(\sigma_{\rm L}(\theta)+\sigma_{\rm R}(\theta)\right),
\efr
subject to the Bethe equations \BEI\ and \BEM\ as a constraint. The result
of the variational problem can be expressed in terms of the 
`excitation energies' for the particles $\epsilon_{\rm L,R}(\theta)$
and the magnons $\xi_p^a(\theta)$, with
$$
\delta
{\cal E}(h)={m_a\over2\pi}\int_{-\infty}^\infty d\theta\left[
\epsilon_{\rm L}^-(\theta)+\epsilon_{\rm R}^-(\theta)\right]\cosh\theta,
\efr
which satisfy the TBA equations:
$$\eqalign{
\epsilon_{\rm L}^+(\theta)+B^{(a)}*\epsilon_{\rm L}^-(\theta)+
C^{(a)}*\epsilon_{\rm
R}^-(\theta)-a_p^{(n)}*\xi_p^{a,-}(\theta)=m_a\cosh\theta-h,\cr
\epsilon_{\rm R}^+(\theta)+B^{(a)}*\epsilon_{\rm R}^-(\theta)+
C^{(a)}*\epsilon_{\rm L}^-(\theta)-a_p^{(n)}*\xi_p^{n-a,-}(\theta)
=m_a\cosh\theta-h,\cr
\xi_p^{b,+}(\theta)+A_{pq}^{(n)}* K_{bc}^{(n)}*\xi_q^{c,-}(\theta)=-\delta_{ba}
a_p^{(n)}*\epsilon_{\rm L}^-(\theta)-\delta_{b,n-a}a_p^{(n)}*
\epsilon_{\rm R}^-(\theta).\cr}
\nfr{TBAE}
In the above, we have defined
$$
f^\pm(\theta)=\cases{f(\theta) &$f(\theta){>\atop<}0$\cr
0&otherwise.\cr}
\efr
The above TBA equations can be drastically simplified because
$a_p^{(n)}(\theta)$ is a positive kernel for all $\theta$; hence the
solution for the magnon terms is simply
$$
\xi_p^b(\theta)=-\delta_{ba}a_p^{(n)}*\epsilon_{\rm
L}^--\delta_{b,n-a}a_p^{(n)}*\epsilon_{\rm R}^-(\theta),
\efr
with the consequence that 
$\xi_p^{b,-}(\theta)=0$. After taking this into account, it is easy to
see that the remaining equations are symmetric in L and R; hence
the solution to \TBAE\ clearly
has $\epsilon_{\rm L}(\theta)=\epsilon_{\rm
R}(\theta)\equiv\epsilon(\theta)$. So finally we are left with a single
integral equation
$$
\epsilon^+(\theta)+R*\epsilon^-(\theta)=m_a\cosh\theta-h,
\nfr{FTBA}
and the change in the ground-state energy density is
$$
\delta {\cal E}(h)={m_a\over\pi}\int_{-\infty}^\infty
d\theta\epsilon^-(\theta)\cosh\theta.
\efr
The Fourier transform of the kernel
$R(\theta)=B^{(a)}(\theta)+C^{(a)}(\theta)$ is explicitly
$$
\hat R(x)={2e^{\pi x/2}\sinh(\pi ax/n)\sinh(\pi(n-a)x/n)\sinh(\pi x/2)
\over\sinh^2(\pi x)}.
\efr

>From the single integral equation we can use the results of
[\Ref{FNW},\Ref{BNNW}] (based on the original work of [\Ref{JNW}]) to
develop an expansion for $\delta {\cal E}(h)$ in the 
asymptotic regime $h\gg m$.
The nature of the solution depends upon whether or not $\hat R(x)$
vanishes at the origin. In the present case we see that $\hat R(0)=0$
which implies that $\delta {\cal E}(h)$ for this model 
is therefore of the
type encountered in the bosonic sigma models, like the principal
chiral models [\Ref{BNNW}], rather than the fermionic models. In other
words it has an expansion which precisely matches the perturbative
result with a classical term of ${\cal O}(1/g)$ present, so that $a_0
\neq 0$.

To find the first few terms in the
expansion of the solution one has to write the Fourier transform of the 
kernel in the form $1/(G_+(x)G_-(x))$ where $G_\pm(x)$ are 
analytic in the upper (lower) half planes with
$G_-(x)=G_+(-x)$. This determines uniquely
$$\eqalign{
G_+(i\xi)=&{n\over\sqrt{\pi a(n-a)\xi}}
{\Gamma(1+a\xi/n)\Gamma(1+(n-a)\xi/n)\Gamma(1+\xi/2)\over
\Gamma^2(1+\xi)}\cr
&\times\exp\left({\xi\over2}\ln\xi+\xi\left(-{1\over2}-{a\over
n}\ln{a\over n}-{n-a\over n}\ln{n-a\over
n}-{1\over2}\ln{1\over2}\right)\right).\cr}
\efr
Following the discussion in [\Ref{BNNW}], if
$G_+(i\xi)$ has an expansion for small $\xi$ like
$$
G_+(i\xi)={k\over\sqrt\xi}e^{-a\xi\ln\xi}\left(1-b\xi+{\cal
O}(\xi^2)\right),
\nfr{EXG}
then the first few terms of the ground-state energy for $h\gg m$ are given by
$$\eqalign{
\delta{\cal E}=&-{h^2k^2\over2}\left[\ln{h\over m_a}
+\ln\left({\sqrt{2\pi}ke^{-b}\over G_+(i)}\right)-1+ 
a(\gamma_{\rm E}-1+\ln8)\right.\cr
&\qquad\qquad\qquad\left.
+(a+\half)\ln\ln{h\over m_a}+
{\cal O}\left({\ln\ln(h/m_a)\over\ln(h/m_a)}\right)
\right].\cr}
\efr
Our kernel does indeed have an expansion of the form \EXG\ with
$$
k={n\over\sqrt{\pi a(n-a)}},\quad
a=-{1\over2},\quad{\sqrt{2\pi}ke^{-b}\over
G_+(i)}={2^{3/2}n^2e^{\gamma_{\rm E}/2}\over\pi a(n-a)}
\sin\left({\pi a\over n}\right), 
\efr
and so the first few terms in the ground-state energy are
$$
\delta{\cal E}(h)=-{h^2n^2\over2\pi a(n-a)}\left[\ln{h\over m}
+\ln\left({n^2\sin(\pi/n)\over\pi a(n-a)}\right)-{1\over2}+
{\cal O}\left({\ln\ln(h/m)\over\ln(h/m)}\right)
\right],
\nfr{FESM}
where we have used the mass formula \MASS\ to relate $m_a$ to the mass of
the lightest multiplet, namely $m$.

\chapter{Comparison and Conclusions}

Comparing \FESM\ with \FEPT , we find that the results from the TBA
calculation and the perturbative calculation are in complete
agreement for each of the charges defined by \WEIGHTS .
It is important to realize that the results of the calcualtions for
these different charges are logically independent; each of them probes
the S-matrix in a slightly different way, and we have shown that they 
all correctly reproduce the universal beta-function coefficients 
written in (3.2).
We take this as very strong evidence that our conjectured S-matrix 
does indeed provide the correct description of 
the supersymmetric SU($n$) principal chiral sigma model.
The comparison also leads to the exact result for the mass-gap:
$$
{m\over\Lambda_{\overline{\rm
MS}}}={n\over\pi}\cdot\sin\left({\pi\over n}\right).
\nfr{MG}
Once again, it is non-trivial that the same result is obtained for
each of the charges in \WEIGHTS .

The question of CDD ambiguities deserves special attention. 
As we have said previously, in writing down any S-matrix proposed on
the basis of symmetries and general axioms, we always have the freedom 
to multiply by CDD factors which respect all the basic principles
automatically, which do not
introduce any new poles onto the physical strip, and which satisfy the
bootstrap equations. In our case this corresponds to multiplying the
S-matrix elements $S_{ab}(\theta)$ by products of factors of the form
\CDD. More precisely, suppose we introduce such factors with parameters
$\alpha_j$ for LL and RR scattering and parameters $\beta_k$ for LR scattering.
The effect of these is to modify the
kernels $B^{(a)}(\theta)$ and $C^{(a)}(\theta)$, and hence the kernel
$R(\theta)$:
$$
\hat R(x)\rightarrow\hat R(x)
-{\hat A_{aa}^{(n)}(x)\over\cosh(\pi x/n)}\sum_j\cosh(\pi(1-\alpha_j)x/n)
-{\hat A_{a,n-a}^{(n)}(x)\over\cosh(\pi x/n)}\sum_k
\cosh(\pi(1-\beta_k)x/n).
\efr
But from (4.6), we see that the modified kernel no longer has $\hat
R(0)=0$ and the
agreement with perturbation theory is therefore destroyed, by virtue
of the remarks made in section 3.
Modification by CDD factors in particular is therefore ruled out if we
are to maintain consistency with the lagrangian.
Of course one could always introduce additional CDD factors
which give extra poles on the physical strip, but in that case the
resulting S-matrix would require the existence of new states.

For the case of SU(2), our result provides the solution of the
supersymmetric O(4) sigma model. As suspected in [\Ref{SW}], the vector
particle of the theory is no longer stable and the spectrum just
consists of the kinks transforming in the spinor and anti-spinor
representations (that is $(1/2,0)\oplus(0,1/2)$ of SU(2)$\times$SU(2)).
It is interesting that the analogous spectrum which we have proposed
for the general super SU($n$) model---and which we have checked by our
calcultions---is markedly different from the bosonic case, with
particles transforming non-trivially under the left-handed or
right-handed symmetry groups, but not under both.
It would be intersting to find some semi-classical
understanding of this. 

Finally, we must return to an important assumption which we made 
and whose validity we have not demonstrated directly,
but which is certainly borne out by our final results.
We have not shown explicitly that our S-matrix for the super SU($n$) PCM 
is invariant under $N=1$ supersymmetry. However, we do know
that it commutes with conserved spin-1/2 charges, it is only their
algebra which has not been directly established for general $n$, 
and in fact this algebra {\it has\/} been checked explicitly for $n=2$
in [\Ref{BL}] and partially by us for $n=3$.
It would be interesting and worthwhile to show explicitly how 
$N=1$ supersymmetry is realized on the SU($n$) RSOS kink S-matrices at
level $n$; in a sense this would relate the construction to the
general scheme for supersymmetric scattering laid out in
[\Ref{SCHOU}].
There are similar---although perhaps not identical---issues 
concerning the $N=2$ S-matrices proposed in [\Ref{LNW}] and to our
knowledge these points are still unresolved.
We believe that the final agreement of our S-matrix with a {\it
supersymmetric\/} lagrangian provides compelling evidence 
that our assumption about supersymmetry is correct; we also
hope to examine this question in more detail in the future.
\sjump

JME acknowledges helpful conservations with E. Abdalla, 
N. MacKay, G. Takacs and G.M.T. Watts.
JME was supported at CERN by a Fellowship from the EU Human Capital
and Mobility programme. 
TJH and JME are both currently supported by PPARC Advanced Fellowships.
\vfill \eject

\references

\beginref
\Rref{FNW}{P. Forg\'acs, F. Niedermayer and P. Weisz, Nucl. Phys. {\bf
B367} (1991) 123}
\Rref{HN}{P. Hasenfratz and F. Niedermayer, Phys. Lett. {\bf B245}
(1990) 529}
\Rref{BNNW}{J. Balog, S. Naik, F. Niedermayer and P. Weisz, Phys. Rev. Lett.
{\bf69} (1992) 873\newline
S. Naik, Nucl. Phys. {\bf B} (Proc. Suppl.) {\bf30} (1993) 232}
\Rref{HMN}{P. Hasenfratz, M. Maggiore and F. Niedermayer, Phys. Lett.
{\bf B245} (1990) 522}
\Rref{ZZ}{A.B. Zamolodchikov and Al. B. Zamolodchikov, Ann. Phys.
{\bf120} (1979) 253}
\Rref{TH3}{T.J. Hollowood, Phys. Lett. {\bf B329} (1994) 450}
\Rref{TBA}{E.H. Lieb and W. Liniger, Phys. Rev. {\bf130} (1963) 1605\newline
Al.B. Zamolodchikov, Nucl. Phys. {\bf B342} (1990) 695}
\Rref{PW}{A. Polyakov and P.B. Wiegmann, Phys. Lett. {\bf B131} (1983)
121}
\Rref{W}{P.B. Wiegmann, Phys. Lett. {\bf B141} (1984) 217}
\Rref{AHN2}{C. Ahn, Nucl. Phys. {\bf B354} (1990) 57}
\Rref{ABL}{C. Ahn, D. Bernard and A. LeClair, Nucl. Phys. {\bf B346}
(1990) 409}
\Rref{LNW}{A. LeClair, D. Nemeschansky and N.P. Warner, Nucl. Phys.
{\bf B390} (1993) 653}
\Rref{CGN}{P. Forg\'acs, S. Naik and F. Niedermayer, 
Phys. Lett. {\bf B283} (1992) 282}
\Rref{AFM}{L. Alvarez-Gaum\'e, D.Z. Freedman and S.K. Mukhi, Ann.
Phys. {\bf134} (1981) 85}
\Rref{SW}{R. Shankar and E. Witten, Phys. Rev. {\bf D17} (1978) 2134}
\Rref{GNK}{R. Shankar and E. Witten, Nucl. Phys. {\bf B141} (1978)
349; E. Witten, Nucl. Phys. {\bf B142} (1978) 285}
\Rref{SCHOU}{K. Schoutens, Nucl. Phys. {\bf B344} (1990) 665}
\Rref{EH}{J.M. Evans and T.J. Hollowood, Nucl. Phys. {\bf B438} (1995)
469}
\Rref{EH1}{J.M. Evans and T.J. Hollowood, Phys. Lett. {\bf B343}
(1995) 189; Phys. Lett. {\bf B343} (1995) 198}
\Rref{WIT}{E. Witten, Phys. Rev. {\bf D16} (1977) 2991}
\Rref{AFL}{E. Abdalla, M. Forger and A. Lima Santos, Nucl. Phys. {\bf
B256} (1985) 145}
\Rref{AAL}{E. Abdalla, M. Abdalla and A. Lima-Santos, Phys. Lett. {\bf
B140} (1984) 71}
\Rref{AAR}{
E. Abdalla, M.C.B. Abdalla and K.D. Rothe, `Non-perturbative methods
in 2 dimensional quantum field theory' (World-Scientific; 1991)}
\Rref{TH1}{T.J. Hollowood, Int. Journ. Mod. Phys. {\bf A8} (1993) 947}
\Rref{TH2}{T.J. Hollowood, Nucl. Phys. {\bf B414} (1994) 404}
\Rref{D1}{G. Delius, Nucl. Phys. {\bf B451} (1995) 445}
\Rref{EH3}{J.M. Evans and T.J. Hollowood, article in preparation}
\Rref{TH4}{T.J. Hollowood, Phys. Lett. {\bf B320} (1994) 43}
\Rref{RM}{F. Ravanini, Phys. Lett. {\bf B282} (1992) 73\newline
M.J. Martins, Phys. Lett. {\bf B277} (1992) 301}
\Rref{JNW}{G. Japaridze, A. Nersesyan and P. Wiegmann,
Nucl. Phys. {\bf B230} (1984) 511}
\Rref{HM}{T.J. Hollowood and E. Mavrikis, in preparation}
\Rref{Col}{S. Coleman, Commun. Math. Phys. {\bf 31} (1973) 259}
\Rref{QG}{G. Felder and A. LeClair, Int. Journ. Mod. Phys. {\bf A7}
Suppl. {\bf 1A} (1992) 239; \newline
D. Bernard and A. LeClair, Commun. Math. Phys. {\bf 142} (1991) 99} 
\Rref{HAS}{M. Hasenbusch and S. Meyer, Phys. Rev. Lett. {\bf68} (1992)
435}
\Rref{HH}{M. Hasenbusch and R.R. Horgan, `{\sl Tests of the continuum
limit for the SO(4) principal chiral model and the prediction for
Lambda(MS)\/}', {\tt hep-lat/951104}}
\Rref{BL}{D. Bernard and A. LeClair, Nucl. Phys. {\bf B340} (1990)
721}
\Rref{ORW}{E. Ogievetsky, N. Reshetikhin and P. Wiegmann,
Nucl. Phys. {\bf B280} (1987) 45}
\Rref{L1}{A. LeClair, Phys. Lett. {\bf B230} (1989) 103}
\Rref{EH4}{J.M. Evans and T.J. Hollowood, Nucl. Phys. {\bf B} (Proc. Suppl.)
{\bf 45A} (1996) 103}
\endref
\ciao

\bibitem{HMN}{P. Hasenfratz, M. Maggiore and F. Niedermayer, Phys. Lett.
{\bf B245} (1990) 522}
\bibitem{HN}{P. Hasenfratz and F. Niedermayer, Phys. Lett. {\bf B245}
(1990) 529}
\bibitem{FNW}{P. Forg\'acs, F. Niedermayer and P. Weisz, Nucl. Phys. {\bf
B367} (1991) 123}
\bibitem{CGN}{P. Forg\'acs, S. Naik and F. Niedermayer, 
Phys. Lett. {\bf B283} (1992) 282}
\bibitem{BNNW}{J. Balog, S. Naik, F. Niedermayer and P. Weisz, Phys. Rev. Lett.
{\bf69} (1992) 873\newline
S. Naik, Nucl. Phys. {\bf B} (Proc. Suppl.) {\bf30} (1993) 232}
\bibitem{THIII}{T.J. Hollowood, Phys. Lett. {\bf B329} (1994) 450}

\bibitem{ZZ}{A.B. Zamolodchikov and Al.B. Zamolodchikov, Ann. Phys.
{\bf120} (1979) 253}
\bibitem{SMGN}{A.B. Zamolodchikov and Al.B. Zamolodchikov, Nucl. Phys.
{\bf B133} (1978) 525\newline
M. Karowski and H.J. Thun, Nucl. Phys. {\bf B190} (1981)
61}
\bibitem{SMCGN}{B. Berg and P. Weisz, Nucl. Phys. {\bf B146} (1979)
205\newline
V. Kurak and J.A. Swieca, Phys. Lett. {\bf B82} (1979) 289}

\bibitem{SMCPN}{R. K\"oberle and V. Kurak, Phys. Rev. {\bf D36} (1987) 627}
\bibitem{AL}{E. Abdalla and A. Lima-Santos, Phys. Rev. {\bf D29}
(1984) 1851}
\bibitem{EH3}{J.M. Evans and T.J. Hollowood, review in preparation}

\bibitem{PW}{A. Polyakov and P.B. Wiegmann, Phys. Lett. {\bf B131} (1983)
121}
\bibitem{W}{P.B. Wiegmann, Phys. Lett. {\bf B141} (1984) 217}
\bibitem{TBA}{E.H. Lieb and W. Liniger, Phys. Rev. {\bf130} (1963) 1605\newline
Al.B. Zamolodchikov, Nucl. Phys. {\bf B342} (1990) 695}

\bibitem{TH2}{T.J. Hollowood, Phys. Lett. {\bf B320} (1994) 43}

\bibitem{AFL}{}
\bibitem{AAG}{E. Abdalla, M.C.B. Abdalla and M. Gomes, Phys. Rev. {\bf
D25} (1982) 452; Phys. Rev. {\bf D27} (1983) 825}
\bibitem

\bibitem{BCR}{P. Biscari, M. Campostrini and P. Rossi, Phys. Lett.
{\bf B242} (1990) 225}
\bibitem{FNWII}{P. Forg\'acs, F. Niedermayer and P. Weisz, Nucl. Phys. {\bf
B367} (1991) 144}
\bibitem{JAG}{J.A. Gracey, Phys. Lett. {\bf B298} (1993) 116
\newline
M. Campostrini and P.Rossi, Int. J. Mod. Phys. {\bf A7} 
(1992) 3265}
\bibitem{MC}{M. Ciuchini, Phys. Lett. {\bf B306} (1993) 59}

\bibitem{INST}{E. Witten, Nucl.~Phys.~{\bf B149} (1979) 285}
\bibitem{DDL2}{A. D'Adda, P. Di Vecchia and M. L\"uscher, 
Nucl.~Phys.~{\bf B152} (1979) 125}
\bibitem{SHOU}{K. Schoutens, Nucl. Phys. {\bf B344} (1990) 665}
\bibitem{FI}{P. Fendley and K. Intriligator, Nucl. Phys. {\bf B380}
(1992) 265}
\bibitem{CV}{S. Cecotti and C. Vafa, Phys.~Rev.~Lett.~{\bf 68} (1992) 903;
Commun.~Math.~Phys.~{\bf 158} (1993) 569}

\bibitem{HAS}{M. Hasenbusch and S. Mayer, Phys. Rev. Lett. {\bf68}
(1992) 435}
